\documentclass[superscriptaddress,twocolumn, pra]{revtex4}
\usepackage{amsmath}
\usepackage{amssymb}
\usepackage{dsfont}
\usepackage{color}
\usepackage{float}
\usepackage{hyperref}
\usepackage{graphicx}
\usepackage{sidecap}
\usepackage{soul}

\emergencystretch=\maxdimen
\usepackage{xcolor}
\hypersetup{
  colorlinks   = true, 
  urlcolor     = black, 
  linkcolor    = black, 
  citecolor   = black 
}

\begin{document}
\title{Experimental observation of anomalous topological edge modes \\ in a slowly-driven photonic lattice}

\author{Sebabrata~Mukherjee\footnote{These authors contributed equally in this work.}}
\email{snm32@hw.ac.uk}
\affiliation{Institute of Photonics and Quantum Sciences, School of Engineering $\&$ Physical Sciences, Heriot-Watt University, Edinburgh, EH14 4AS, United Kingdom}
\author{Alexander~Spracklen$^{*}$}
\email{aos32@hw.ac.uk}
\affiliation{Institute of Photonics and Quantum Sciences, School of Engineering $\&$ Physical Sciences, Heriot-Watt University, Edinburgh, EH14 4AS, United Kingdom}
\author{Manuel~Valiente}
\affiliation{Institute of Photonics and Quantum Sciences, School of Engineering $\&$ Physical Sciences, Heriot-Watt University, Edinburgh, EH14 4AS, United Kingdom}
\author{Erika~Andersson}
\affiliation{Institute of Photonics and Quantum Sciences, School of Engineering $\&$ Physical Sciences, Heriot-Watt University, Edinburgh, EH14 4AS, United Kingdom}
\author{Patrik \"Ohberg}
\affiliation{Institute of Photonics and Quantum Sciences, School of Engineering $\&$ Physical Sciences, Heriot-Watt University, Edinburgh, EH14 4AS, United Kingdom}
\author{Nathan~Goldman}
\affiliation{Center for Nonlinear Phenomena and Complex Systems, Universit\'e Libre de Bruxelles, CP 231, Campus Plaine, B-1050 Brussels, Belgium} 
\author{Robert R. Thomson}
\affiliation{Institute of Photonics and Quantum Sciences, School of Engineering $\&$ Physical Sciences, Heriot-Watt University, Edinburgh, EH14 4AS, United Kingdom}

\begin{abstract}
The discovery of the quantised Hall effect, and its subsequent topological explanation, demonstrated the important role topology can play  in determining the properties of quantum systems~\cite{TKNN,Hasan,Zhang}. This realisation led to the development of topological band theory, where, in addition to band index and quasimomentum, Bloch bands are also characterised by a set of topological invariants.  This topological theory can be readily extended to periodically-driven systems. In the limit of fast driving, the topology of the system can still be captured by the topological invariants used to describe static systems~\cite{kitagawa2010topological,lindner2011floquet}. In the limit of slow driving, however, situations can arise where standard topological invariants are zero, but yet, topologically protected edge modes are still observed~\cite{kitagawa2010topological, rudner2013anomalous, nathan2015topological, Roy, Platero}. These ``anomalous" topological edge modes have no static analogue, and are associated with a distinct topological invariant, which takes into account the full time-evolution over a driving period. Here we demonstrate the first experimental observation of such anomalous topological edge modes in an ultrafast-laser-inscribed photonic lattice. This inscription technique allows one to address each bond of a lattice independently and dynamically, generating a rich band structure with  robust anomalous chiral edge modes and the potential for perfectly localised bulk states. 
\end{abstract}
\maketitle

\begin{figure}[t!]
\includegraphics[width=8.6cm]{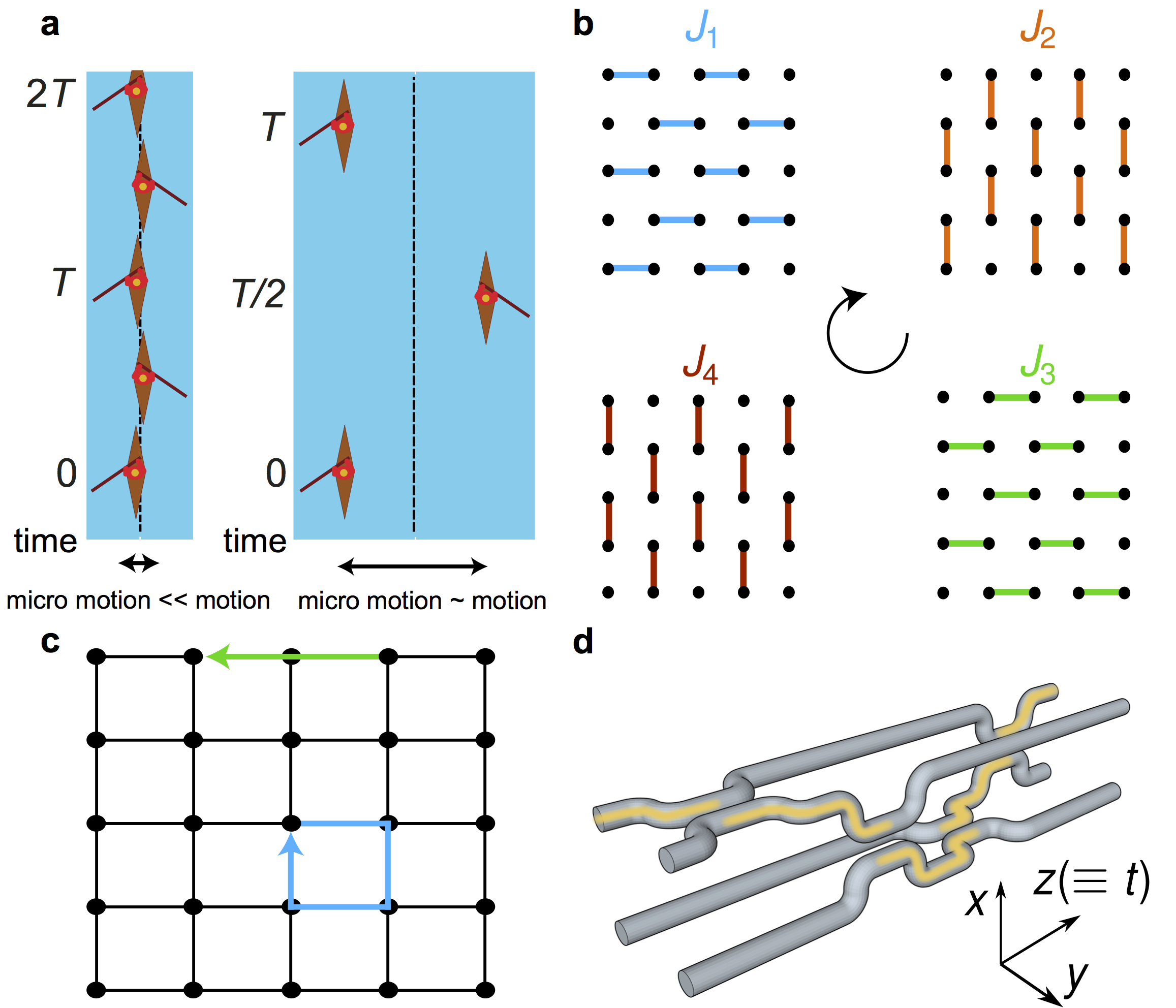}
\caption{{\bf Slowly-driven systems and photonic implementation.} (a) Sketch illustrating the high-frequency (left) and the low-frequency (right) regimes of  a simple driven system, here represented by a rower. In the high-frequency regime, the micro-motion typically only slightly affects the motion: the time-evolution is well captured by  time-averaged dynamics, which can be probed stroboscopically at discrete times $\Delta_t\!=\! T \!\times\! \text{integer}$. In the low-frequency regime, the micro-motion can significantly affect the dynamics, and the time-evolution is no longer well described by the time-averaged Hamiltonian only.
(b) The four different bonds present in the lattice (with coupling constants $J_{1,2,3,4}$) and the cyclic driving protocol employed. (c) In the simplest case where all bond strengths are equal, $JT/4=\pi/2$, chiral edge modes arise whilst the evolution operator associated with the bulk is identity. (d) Sketch illustrating how, using laser inscription, different pairs of waveguides can be moved together to turn on a coupling and then apart to switch it off. This flexibility allows for a realisation of the driving protocol shown in (b). In waveguide arrays the propagation direction $z$ plays a role analogous to time~\protect\cite{garanovich2012light}. }
\label{fig1}
\end{figure}

Subjecting a system to time-periodic modulations constitutes a powerful method to engineer band structures with non-trivial topological properties
~\cite{sorensen2005fractional, kitagawa2010topological, lindner2011floquet, hauke2012non, goldman2014periodically, cayssol2013floquet}, as recently demonstrated in cold-atom experiments
~\cite{jotzu2014experimental, aidelsburger2015measuring} and photonics
~\cite{rechtsman2013photonic}. In this \emph{Floquet-engineering} approach, a static system, described by a Hamiltonian $\hat H_0$, is driven periodically in time by a modulation $\hat V (t)\!=\!\hat V(t+T)$, whose frequency will be denoted $\omega=2 \pi/T$. The time-evolution operator over an arbitrary long duration $\Delta_t \!=\! t_{\text{f}} - t_{\text{0}}$ then takes the general form
~\cite{goldman2014periodically,Fishman}
\begin{equation}
\hat U (t_0\!\rightarrow\!t_{\text{f}})=e^{-i \hat K (t_{\text{f}})} e^{-i \Delta_t \hat H_{\text{eff}}} e^{i \hat K (t_0)}.\label{time-evolution}
\end{equation}
Here, the time-independent (effective) Hamiltonian $\hat H_{\text{eff}}$ describes the time-averaged dynamics over the duration $\Delta_t$, which can be  isolated by probing the system stroboscopically at discrete times $\Delta_t\!=\! T \!\times\! \text{integer}$. This effective Hamiltonian stems from a rich interplay between the static system $\hat H_0$ and the time-modulation $\hat V (t)$~
\cite{goldman2014periodically, cayssol2013floquet}; its band structure (the quasienergies or ``Floquet spectrum") can feature topological properties, such as non-zero Chern numbers and chiral edge modes 
\cite{sorensen2005fractional, kitagawa2010topological, lindner2011floquet, hauke2012non, goldman2014periodically, cayssol2013floquet}, in direct analogy with non-driven systems. The time-evolution operator [Eq.~\eqref{time-evolution}] also includes the  micro-motion operator $\hat K (t_{\text{f}})$, which describes the dynamics that takes place within each period of the driving, i.e.~$\Delta_t\!\ne\! T \!\times\! \text{integer}$. In the high-frequency regime of the driving ($\omega \rightarrow \infty$), observables are generally only slightly affected by the micro-motion, Fig.~\ref{fig1}(a). In this regime, the topological properties of the driven system are then entirely captured by the effective Hamiltonian $\hat H_{\text{eff}}$. The observation of topological edge modes can be directly related to the Chern numbers associated with the effective band structure~\cite{kitagawa2010topological, lindner2011floquet}. Such edge modes were directly visualized in photonic realizations of Floquet band structures~\cite{rechtsman2013photonic} (see also Refs.~\cite{Hafezi2011Robust,Hafezi2013Imaging,Hafezi2014synthetic,soljacicChern} for other photonic implementations of standard topological edge modes). Importantly, this is no longer the case in the low-frequency regime, where $\hbar \omega$ becomes comparable to the typical energy scales of the system (e.g.~the band-width of the effective band structure~\cite{kitagawa2010topological, rudner2013anomalous, nathan2015topological, Roy, Platero, Smith1, Smith2}). In this regime, the micro-motion strongly affects the time-evolution of observables [Fig.~\ref{fig1}(a)], and the topological properties of the system can no longer be simply related to the effective Hamiltonian only. In particular, the presence of topological edge modes is now entirely ruled by a distinct topological invariant, a winding number $W$ that takes into account the micro-motion
~\cite{rudner2013anomalous, nathan2015topological}. In the following, topological edge modes that emerge in the low-frequency regime, due to the micro-motion, will be referred to as ``anomalous" topological edge modes, as proposed in Ref.~\cite{rudner2013anomalous}. Since these topological states are captured by the winding numbers $W$, and not by the Chern numbers of the effective Hamiltonian, they constitute ``genuine Floquet" topological modes. This intriguing regime of periodically-driven systems has recently been investigated in a diverse range of experimental platforms: topologically protected bound states associated with non-trivial $W$ were first demonstrated in a one-dimensional photonic setup realizing a discrete time quantum walk~\cite{Kitagawabound}; more recently Ref.~\cite{soljacicanomalous} reported on the observation of anomalous edge modes in a designer surface plasmon platform. Finally, Ref.~\cite{microwave} considered a Thouless-pump approach to measure $W$ in a one-dimensional microwave network.

The theory works in Ref.~\cite{kitagawa2010topological,rudner2013anomalous} introduced conceptually simple models that exhibited such intriguing Floquet topological structures. 
The model in Ref.~\cite{rudner2013anomalous} is a square lattice with nearest-neighbour couplings which are engineered so that the couplings between a lattice site and its four nearest neighbours are independently controllable. These four couplings, denoted $J_1$ through $J_4$, are then varied in a spatially homogeneous and time-periodic manner so that any lattice site is, at any given moment,  coupled to only one of its nearest neighbours, see Fig.~\ref{fig1} (b). The simplest demonstration of this model is when the driving period $T$ is split into four equal steps, and $T$ is selected, such that a particle that is initially localised on a certain site will be completely transferred to the neighbouring site after a time $T/4$. 
Therefore, considering for now a system without edges, this means that after one complete period, any initial state is exactly reproduced, i.e. the propagator, $\hat{U}(T)$, is the identity matrix [Fig.~\ref{fig1} (c)]. As a corollary, the effective Hamiltonian in Eq.~\eqref{time-evolution} is the zero matrix and the Floquet spectrum consists of two degenerate flat bands at zero energy. The bulk is completely localised, and the Chern numbers associated with the effective Hamiltonian are necessarily trivial. In a geometry with edges, however, it is found that there are chiral propagating edge modes that are localised along the edge. These occur because a particle launched at an edge cannot get back to its initial position and ends up moving one unit cell along the edge [Fig.~\ref{fig1} (c)], in direct analogy with the skipped cyclotron orbits of quantum Hall systems. 

As previously mentioned, the topological properties of slowly-driven systems are well captured by a winding number that takes the full time evolution into account, including the micro-motion~\cite{rudner2013anomalous, nathan2015topological}. For the two-band model introduced above, the time evolution operator at time $t$, where $0\leq t<T$,  may be written as 
\begin{equation}
\hat U(\textbf{k},t)=\sum_{n=1,2} \hat P_n(\textbf{k},t) e^{-i\phi_{n}(\textbf{k},t)},
\label{Proj}
\end{equation}
where $\hat P_n(\textbf{k},t)$ and $e^{-i\phi_{n}(\textbf{k},t)}$ are the projectors and eigenvalues of $\hat U(\textbf{k},t)$, respectively.  The instantaneous ``energies", $\phi_{n}/T$, are defined modulo $2\pi/T$, and we define the corresponding ``Floquet-Brillouin zone" to be in the range $[-\pi/T,\!\quad\!\pi/T]$. In a driven system, there are two types of degeneracy that can occur within a driving period: inter-zone degeneracy, where $\phi_{1}(\textbf{k},t)=\phi_{2}(\textbf{k},t)$, and degeneracy through the zone edge where $\phi_{1}(\textbf{k},t)=-\pi$ and $\phi_{2}(\textbf{k},t)=\pi$. It is the existence of this latter type of degeneracy that profoundly alters the topological picture of driven systems. Zone-edge singularities (ZES) allow the Chern number of the bands to change without the inter-zone gap closing. Consequently, these zone-edge degeneracies can lead to violations of the static bulk-edge correspondence~\cite{bulkedge,bulkedge1}, {so that} 
the Chern number of both Floquet bands are zero but yet protected edge modes are still present~\cite{kitagawa2010topological}. The adequate topological characterisation of driven systems is captured by a winding number which includes the changes in Chern numbers that occur through the zone edge, within a period of the driving. In an  two-band driven system, there can be at most two bandgaps, and for each of these bandgaps, there is a winding number $W_m$, which has the form~\cite{nathan2015topological}    
\begin{equation}
W_m=\sum_{n=1}^{m} C_{n} -\sum_i q_i^{\text{ZES}}, \quad m=1,2,
\label{Winding}
\end{equation}
where $C_n$ is the Chern number of the $n$-{th} band at time $T$, and where  $q_i$ corresponds to the change in Chern number of the lowest band that occurs in the $i$-{th} zone edge degeneracy.
This winding number directly gives the number, $n_{\text{edge}}(m)$, of topologically protected chiral edge modes present in the $m$-{th} gap. This ``driven-system" bulk-edge correspondence can be shown to have the form~\cite{nathan2015topological}
\begin{equation}
n_{edge}(m)=W_m=\sum_{n=1}^{m} C_{n} -\sum_i q_i^{\text{ZES}}.
\label{EdgeNum}
\end{equation}
The first term in the second equality of Eq.~\eqref{EdgeNum} is the term that is found in static systems, whilst the second applies only to driven systems. It is this term that is the source of the anomalous edge modes as it allows the number of edge modes to be non-zero even if the standard topological invariants, i.e.~the Chern numbers $C_n$, are zero for all of the bands.

\begin{figure}[t!]
\includegraphics[width=1\linewidth]{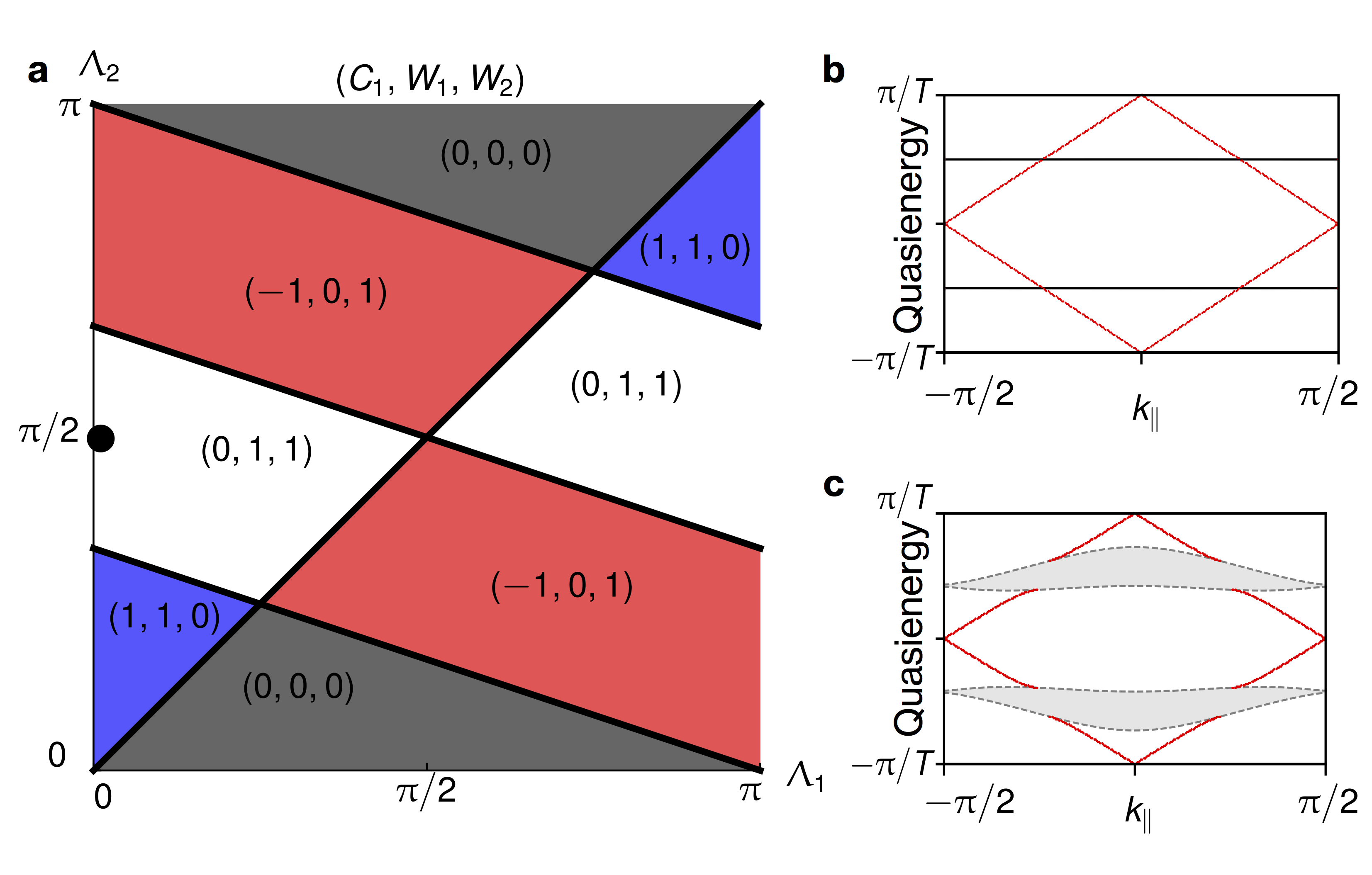}
\caption{{\bf Topological phase diagram and band structures.} (a) Phase diagram for a generalisation of the model in Fig.~$\ref{fig1}(b)$, which allows the bond $J_1$ to be of a different strength to the bonds $J_{2-4}\!=\!\tilde{J}$. Here $\Lambda_1\!=\!J_1T/4$ and $\Lambda_2\!=\!\tilde{J}T/4$. Each phase is described by three topological invariants: the Chern number of the lowest Floquet band  $C_1$, the winding number $W_1$ of the gap centred around zero, and the winding number $W_2$ of the gap centred around $\pi/T$. The white regions indicate where anomalous edge modes can be observed ($C_1\!=\!0, W_{1,2}\!\ne\!0$). (b) Quasienergy spectrum for the parameters indicated by the black dot in (a). Red curves indicate topological-edge-modes dispersions, while black curves correspond to the bulk bands. (c) Quasienergy spectrum for the experimentally achieved parameters. The first Floquet-Brillouin zone is taken to be defined in the range $[-\pi/T,\!\quad\!\pi/T]$.}
\label{fig2}
\end{figure}

\begin{figure*}
 \includegraphics[width=1\linewidth]{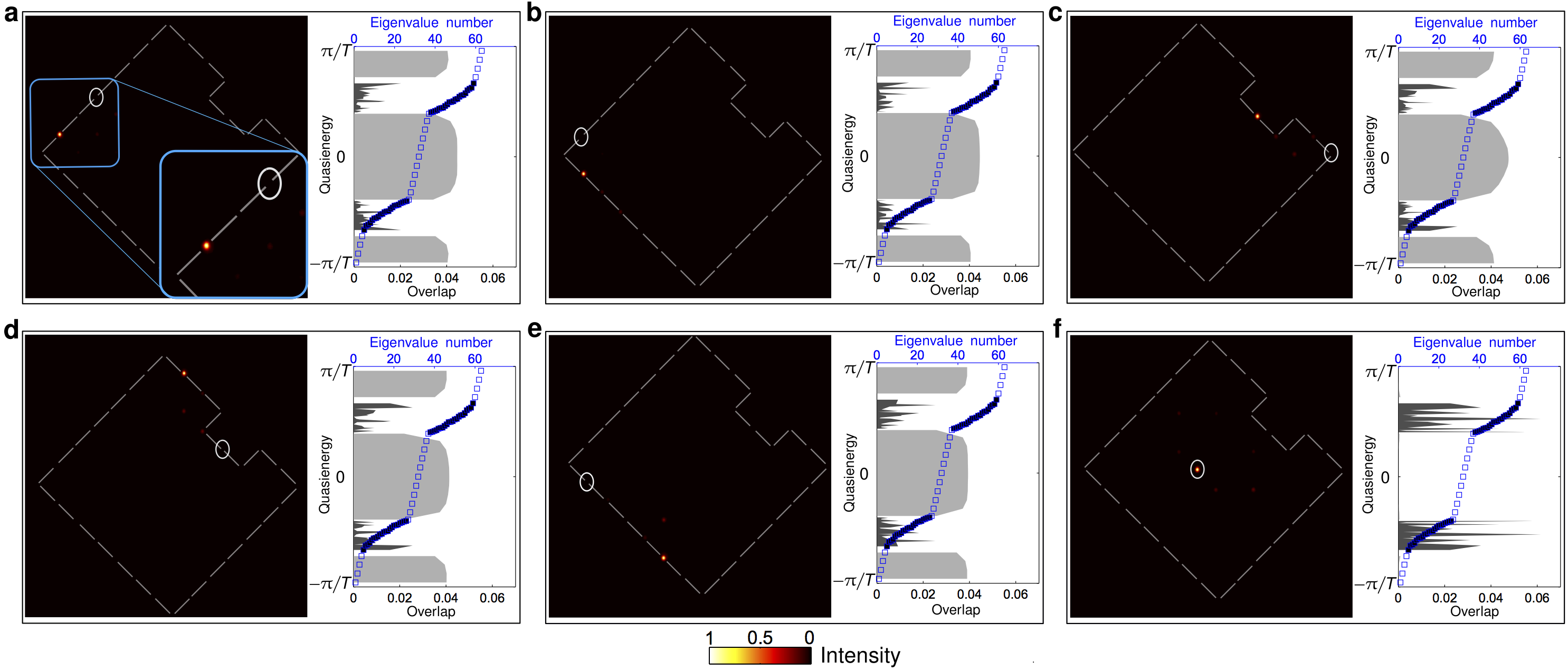}
\caption{{\bf Experimental observation of anomalous topological edge modes.} The left-hand image in each of the panels displays the experimentally measured output intensity distribution when the light is launched at the lattice site indicated by the white circle. The figures display chiral edge modes (a-e) that are not scattered by corners (b) nor defects (c-d) as well as a largely localised bulk state (f). The group velocity of the chiral edge modes along the bottom left edge of the lattice, (e), is twice of that along the top left edge, (a), which stems from the bond $J_1$ having a different strength to the other three bonds. The right-hand image in each panel shows the numerically calculated quasienergy spectrum for the 63-site lattice and the overlap, as a function of this quasienergy, of the different initial states with the  Floquet states. The filled blue squares correspond to the quasienergy of bulk Floquet states whilst the empty blue squares correspond to edge modes. The right-hand panels in (a-e) demonstrate that single-site excitation on the edge of the lattice predominately excites the topologically protected edge modes which thereby yield the robust chiral motion that is observed. Note that the axes of the photonic lattice were rotated by 45$^o$ with respect to the vertical to obtain equal coupling constants along the two axes; see Ref.~\cite{mukherjee2015observation}.}
\label{Exp}
\end{figure*}

In this work, we experimentally implement a variant of the model introduced in Ref.~\cite{rudner2013anomalous}. 
In our model the coupling of the first bond ($J_1$) is different than the following three bonds ($J_{2,3,4}=\!\tilde{J}$), which are chosen so as to satisfy $J_1T/4\!=\!\Lambda_1$ and  $\tilde{J}T/4\!=\!\Lambda_2$.
This gives two controllable experimental  parameters, $\Lambda_1$ and $\Lambda_2$, that can be tuned independently. We find that the interplay between these two parameters produces a rich phase diagram, shown in Fig.~\ref{fig2}~(a), which demonstrates how this simple model can be used to explore many different topological regimes (see Sup.~Mat.). Each topological phase of this driven two-band system is accurately labelled by the winding numbers associated with the two bandgaps; in the phase diagram shown  in Fig.~\ref{fig2}~(a), we have chosen $W_1$ to be the winding number for the bandgap centred around quasienergy zero and $W_2$ for the bandgap centred around $\pi/T$. In addition, we also provide the Chern number of the lowest Floquet band $C_1$, so as to highlight the anomalous regimes where $C_1\!=\!0$ and $W_{1,2}\!\ne\!0$.

For the sake of experimental practicability, we aim to focus on the anomalous edge modes that are predicted for $\Lambda_1=0$ and $\Lambda_2=\pi/2$, as indicated by the black dot in Fig.~\ref{fig2}~(a). In this configuration, the blue bonds in Fig.~\ref{fig1}~(b) are never turned on, while nearly complete transfer of light can occur via the other three bonds.
These parameters are desirable as not only is the resultant system located well within the anomalous regime [Fig.~\ref{fig2}~(a)] but also the associated spectrum consists of two gapped flat bands [Fig.~\ref{fig2}~(b)].
Additionally, the Floquet states have a simple analytical form, and there are robust chiral edges states that can be excited with unit efficiency, through the simple experimental technique of single-site excitation.

The realisation of this model requires a high degree of control over the couplings present in the lattice. Ultrafast-laser-inscribed arrays of optical waveguides with three-dimensional geometry
offer this control, as they allow the coupling between lattice sites to be individually controlled.
In the experiment presented here, a lattice constant of 40~$\mu m$ was chosen, such that the coupling between lattice sites was insignificant over the maximum observable propagation distance. To turn on the coupling between any two waveguides, the inter-waveguide separation is reduced such that the two waveguides propagate together for a 4.5~mm straight section, with a   centre-to-centre waveguide separation of 11~$\mu m$. After this interaction region, the waveguides again separate in a reverse manner, see the sketch in Fig.~\ref{fig1}~(d).
For a particular wavelength, the coupling that occurs between two  synchronously bending waveguides can be shown to be equivalent to that of two straight waveguides with some effective coupling constant  (Sup.~Mat.). 
This effective coupling can be controlled by changing the interaction length and/or the separation of the waveguides; these parameters are fixed upon writing the sample and so do not allow fine-tuning of the coupling in-situ. However, the wavelength of the light used to excite the lattice is a tunable parameter  that can be used to control the effective coupling strength (Sup. Mat.).
To demonstrate the existence of anomalous edge modes, a lattice of size $8\times 8$, consisting of two driving cycles, was fabricated. This lattice also contains a single defect, namely, a missing lattice site on the edge, allowing for the detection of the robustness of edge modes.   

To characterise the bonds present in the lattice, and hence validate our theoretical model, we fabricated five sets of each bond in isolation, inside the same substrate, and measured their behaviour as a function of the wavelength of light (Sup.~Mat.). When characterised with 785~nm light, the mean and standard deviation of $ \Lambda_i$ ($i\!=\!2,3,4$) were found to be $\Lambda_2\!=\!\pi/2 (1.16\pm 0.04)$, $\Lambda_3\!=\!\pi/2 (1.15\pm 0.04)$, $\Lambda_4\!=\!\pi/2 (0.85\pm 0.03)$. For the bond $J_1$, all the light remains in the waveguide excited at the input, which indicates no transfer of light ($\Lambda_1\!=\! 0$). The close proximity of these measured couplings to the aforementioned desired values suggest that anomalous edge modes should be detected for an input wavelength of 785~nm.  Indeed, the quasienergy spectrum corresponding to these experimental values [Fig.~\ref{fig2}(c)] is similar to the ideal case [Fig.~\ref{fig2}(b)]. While a finite dispersion of the bulk now becomes apparent, we note that the velocity of the edge modes remains significantly larger compared to the dispersion of the bulk. It should be noted that the non-zero standard deviations indicate that there will be bond-strength disorder within the lattice, whilst a detailed numerical studies of the experimental results also suggest the additional presence of a phase disorder effect; the latter can be phenomenologically modelled by a small random on-site term. However, we verified that the topological properties of the lattice are unaffected by these sources of disorder (Sup. Mat.).

In order to experimentally demonstrate the presence of anomalous topological edge modes, 785~nm light was launched at multiple locations around the edge of the lattice;  the white circles in Fig.~\ref{Exp} indicate the launch site. If the light is launched at the middle of an edge, as shown in Fig.~\ref{Exp}~(a), then at the output it is observed that the light has moved along the edge with minimal penetration into the bulk. Moreover, if the input position of the beam is moved further down the edge, Fig.~\ref{Exp}~(b), the close proximity of the launch site to the left edge means that the light will encounter the corner of the lattice during its evolution. This corner, as can be observed in the figure, does not cause backscattering but instead the light simply turns the corner and continues to propagate. This robustness is also observed in Fig.~\ref{Exp}~(c-d), which demonstrates the light moving around a missing lattice site without backscattering or penetrating into the bulk. These observations provide evidence for the existence of protected chiral edge modes that can be almost exclusively excited by single-site excitation. These experimental observations are well complemented by a theoretical analysis, which was obtained using the couplings extracted from the bond characterisation data. The right-hand image in each panel of Fig.~\ref{Exp} shows the quasienergy spectrum for the lattice and the overlap, as a function of this quasienergy, of the different initial launch states with the numerically calculated Floquet states. Fig.~\ref{Exp} (a-e) demonstrate how launching in a single edge site leads to almost  ideal excitation of the edge modes with little probability for bulk modes being excited, even when the launch site is close to a defect or corner. The absence of this bulk coupling accounts for the strong confinement of the light to the lattice edge.

The central result of these experimental images is that they show the existence of edge modes propagating in a chiral (unidirectional) manner, and which are not scattered by corners nor defects. The absence of scattering provides strong evidence that there is topological protection for  these edge modes. From the bond-strength measurements, combined with our theoretical analysis of the corresponding model, we demonstrate that these propagating states correspond to anomalous topological edge modes.

The single-site excitation that is used in the experimental setup can also be exploited to probe the different edge modes that are present in the lattice.  Moving the input site onto the bottom left edge, Fig.~\ref{Exp}~(e), shows the light propagating with twice the group velocity to what is observed in  Fig.~\ref{Exp}~(a). This experimentally demonstrates that, when the bond $J_1$ is different to the other three bonds, edge modes with different group velocities can be observed. The exact dispersion relations of the edge modes can be altered by modifying the strength of $J_1$.  The ability to tune the group velocity could present a valuable experimental tool for the future, which could allow, for instance, the investigation of how the non-scattering behaviour of edge modes is modified by the interplay between group velocity and non-linearity.

The  behaviour observed when light is launched in the bulk of the lattice, Fig.~\ref{Exp}~(f),  is markedly different to what occurs on the edge with the input state being almost  exactly reformed at the output facet. This refocusing is due to the beating between the two bulk bands which are separated by a bandgap of approximately $\pi/T$. In an ideal lattice, $\Lambda_2\!=\!\pi/2$, the bulk bands would be dispersionless [Fig.~\ref{fig2}(b)] and so the initial state would continue to reform every two driving cycles. In the experimental lattice, however, deviations of the bonds strengths from the ideal case cause the bands to become dispersive [Fig.~\ref{fig2}(c)], which indicates that the initial state would eventually disperse in lattices of very long propagation lengths. These deviations could be addressed in future work by careful optimisation of the experimental setup and the waveguide bending profile, so as to allow a closer realisation of the ideal model. 

The ability to easily excite edge modes, with almost unit efficiency, and the possible co-existence of chiral edge modes with a dispersionless bulk, are two particularly interesting features of this slowly-driven photonic system. These properties make this lattice a promising platform for investigating topological transport properties in response to perturbations, such as external (engineered) fields, disorder, and particle-particle interaction (as generated by optical non-linearities).

\noindent{\it Acknowledgements}
S.~M. thanks Heriot-Watt University for a James Watt PhD Scholarship. A.~S. acknowledges support from the EPSRC CM-CDT. N.~G. thanks D. T. Tran for discussions, and also the FRS-FNRS Belgium and the BSPO under the PAI project P7/18 DYGEST for support. M.~V. and P.~\"O. acknowledge support from EPSRC  EP/M024636/1. R.~R.~T. gratefully acknowledges funding from the UK Science and Technology Facilities Council (STFC) in the form of an STFC Advanced Fellowship (ST/H005595/1).


\begin{thebibliography}{36}
\expandafter\ifx\csname natexlab\endcsname\relax\def\natexlab#1{#1}\fi
\expandafter\ifx\csname bibnamefont\endcsname\relax
  \def\bibnamefont#1{#1}\fi
\expandafter\ifx\csname bibfnamefont\endcsname\relax
  \def\bibfnamefont#1{#1}\fi
\expandafter\ifx\csname citenamefont\endcsname\relax
  \def\citenamefont#1{#1}\fi
\expandafter\ifx\csname url\endcsname\relax
  \def\url#1{\texttt{#1}}\fi
\expandafter\ifx\csname urlprefix\endcsname\relax\def\urlprefix{URL }\fi
\providecommand{\bibinfo}[2]{#2}
\providecommand{\eprint}[2][]{\url{#2}}

\bibitem[{\citenamefont{Thouless et~al.}(1982)\citenamefont{Thouless, Kohmoto,
  Nightingale, and Den~Nijs}}]{TKNN}
\bibinfo{author}{\bibfnamefont{D.}~\bibnamefont{Thouless}},
  \bibinfo{author}{\bibfnamefont{M.}~\bibnamefont{Kohmoto}},
  \bibinfo{author}{\bibfnamefont{M.}~\bibnamefont{Nightingale}},
  \bibnamefont{and} \bibinfo{author}{\bibfnamefont{M.}~\bibnamefont{Den~Nijs}},
  \bibinfo{journal}{Physical Review Letters} \textbf{\bibinfo{volume}{49}},
  \bibinfo{pages}{405} (\bibinfo{year}{1982}).

\bibitem[{\citenamefont{Hasan and Kane}(2010)}]{Hasan}
\bibinfo{author}{\bibfnamefont{M.~Z.} \bibnamefont{Hasan}} \bibnamefont{and}
  \bibinfo{author}{\bibfnamefont{C.~L.} \bibnamefont{Kane}},
  \bibinfo{journal}{Reviews of Modern Physics} \textbf{\bibinfo{volume}{82}},
  \bibinfo{pages}{3045} (\bibinfo{year}{2010}).

\bibitem[{\citenamefont{Qi and Zhang}(2011)}]{Zhang}
\bibinfo{author}{\bibfnamefont{X.-L.} \bibnamefont{Qi}} \bibnamefont{and}
  \bibinfo{author}{\bibfnamefont{S.-C.} \bibnamefont{Zhang}},
  \bibinfo{journal}{Reviews of Modern Physics} \textbf{\bibinfo{volume}{83}},
  \bibinfo{pages}{1057} (\bibinfo{year}{2011}).

\bibitem[{\citenamefont{Kitagawa et~al.}(2010)\citenamefont{Kitagawa, Berg,
  Rudner, and Demler}}]{kitagawa2010topological}
\bibinfo{author}{\bibfnamefont{T.}~\bibnamefont{Kitagawa}},
  \bibinfo{author}{\bibfnamefont{E.}~\bibnamefont{Berg}},
  \bibinfo{author}{\bibfnamefont{M.}~\bibnamefont{Rudner}}, \bibnamefont{and}
  \bibinfo{author}{\bibfnamefont{E.}~\bibnamefont{Demler}},
  \bibinfo{journal}{Physical Review B} \textbf{\bibinfo{volume}{82}},
  \bibinfo{pages}{235114} (\bibinfo{year}{2010}).

\bibitem[{\citenamefont{Lindner et~al.}(2011)\citenamefont{Lindner, Refael, and
  Galitski}}]{lindner2011floquet}
\bibinfo{author}{\bibfnamefont{N.~H.} \bibnamefont{Lindner}},
  \bibinfo{author}{\bibfnamefont{G.}~\bibnamefont{Refael}}, \bibnamefont{and}
  \bibinfo{author}{\bibfnamefont{V.}~\bibnamefont{Galitski}},
  \bibinfo{journal}{Nature Physics} \textbf{\bibinfo{volume}{7}},
  \bibinfo{pages}{490} (\bibinfo{year}{2011}).

\bibitem[{\citenamefont{Rudner et~al.}(2013)\citenamefont{Rudner, Lindner,
  Berg, and Levin}}]{rudner2013anomalous}
\bibinfo{author}{\bibfnamefont{M.~S.} \bibnamefont{Rudner}},
  \bibinfo{author}{\bibfnamefont{N.~H.} \bibnamefont{Lindner}},
  \bibinfo{author}{\bibfnamefont{E.}~\bibnamefont{Berg}}, \bibnamefont{and}
  \bibinfo{author}{\bibfnamefont{M.}~\bibnamefont{Levin}},
  \bibinfo{journal}{Physical Review X} \textbf{\bibinfo{volume}{3}},
  \bibinfo{pages}{031005} (\bibinfo{year}{2013}).

\bibitem[{\citenamefont{Nathan and Rudner}(2015)}]{nathan2015topological}
\bibinfo{author}{\bibfnamefont{F.}~\bibnamefont{Nathan}} \bibnamefont{and}
  \bibinfo{author}{\bibfnamefont{M.~S.} \bibnamefont{Rudner}},
  \bibinfo{journal}{New Journal of Physics} \textbf{\bibinfo{volume}{17}},
  \bibinfo{pages}{125014} (\bibinfo{year}{2015}).

\bibitem[{\citenamefont{Roy and Harper}(2016)}]{Roy}
\bibinfo{author}{\bibfnamefont{R.}~\bibnamefont{Roy}} \bibnamefont{and}
  \bibinfo{author}{\bibfnamefont{F.}~\bibnamefont{Harper}},
  \bibinfo{journal}{arXiv preprint arXiv:1603.06944}  (\bibinfo{year}{2016}).

\bibitem[{\citenamefont{G{\'o}mez-Le{\'o}n and Platero}(2013)}]{Platero}
\bibinfo{author}{\bibfnamefont{A.}~\bibnamefont{G{\'o}mez-Le{\'o}n}}
  \bibnamefont{and} \bibinfo{author}{\bibfnamefont{G.}~\bibnamefont{Platero}},
  \bibinfo{journal}{Physical Review Letters} \textbf{\bibinfo{volume}{110}},
  \bibinfo{pages}{200403} (\bibinfo{year}{2013}).

\bibitem[{\citenamefont{S{\o}rensen et~al.}(2005)\citenamefont{S{\o}rensen,
  Demler, and Lukin}}]{sorensen2005fractional}
\bibinfo{author}{\bibfnamefont{A.~S.} \bibnamefont{S{\o}rensen}},
  \bibinfo{author}{\bibfnamefont{E.}~\bibnamefont{Demler}}, \bibnamefont{and}
  \bibinfo{author}{\bibfnamefont{M.~D.} \bibnamefont{Lukin}},
  \bibinfo{journal}{Physical Review Letters} \textbf{\bibinfo{volume}{94}},
  \bibinfo{pages}{086803} (\bibinfo{year}{2005}).

\bibitem[{\citenamefont{Hauke et~al.}(2012)\citenamefont{Hauke, Tieleman, Celi,
  {\"O}lschl{\"a}ger, Simonet, Struck, Weinberg, Windpassinger, Sengstock,
  Lewenstein et~al.}}]{hauke2012non}
\bibinfo{author}{\bibfnamefont{P.}~\bibnamefont{Hauke}},
  \bibinfo{author}{\bibfnamefont{O.}~\bibnamefont{Tieleman}},
  \bibinfo{author}{\bibfnamefont{A.}~\bibnamefont{Celi}},
  \bibinfo{author}{\bibfnamefont{C.}~\bibnamefont{{\"O}lschl{\"a}ger}},
  \bibinfo{author}{\bibfnamefont{J.}~\bibnamefont{Simonet}},
  \bibinfo{author}{\bibfnamefont{J.}~\bibnamefont{Struck}},
  \bibinfo{author}{\bibfnamefont{M.}~\bibnamefont{Weinberg}},
  \bibinfo{author}{\bibfnamefont{P.}~\bibnamefont{Windpassinger}},
  \bibinfo{author}{\bibfnamefont{K.}~\bibnamefont{Sengstock}},
  \bibinfo{author}{\bibfnamefont{M.}~\bibnamefont{Lewenstein}},
  \bibnamefont{et~al.}, \bibinfo{journal}{Physical Review Letters}
  \textbf{\bibinfo{volume}{109}}, \bibinfo{pages}{145301}
  (\bibinfo{year}{2012}).

\bibitem[{\citenamefont{Goldman and Dalibard}(2014)}]{goldman2014periodically}
\bibinfo{author}{\bibfnamefont{N.}~\bibnamefont{Goldman}} \bibnamefont{and}
  \bibinfo{author}{\bibfnamefont{J.}~\bibnamefont{Dalibard}},
  \bibinfo{journal}{Physical Review X} \textbf{\bibinfo{volume}{4}},
  \bibinfo{pages}{031027} (\bibinfo{year}{2014}).

\bibitem[{\citenamefont{Cayssol et~al.}(2013)\citenamefont{Cayssol, D{\'o}ra,
  Simon, and Moessner}}]{cayssol2013floquet}
\bibinfo{author}{\bibfnamefont{J.}~\bibnamefont{Cayssol}},
  \bibinfo{author}{\bibfnamefont{B.}~\bibnamefont{D{\'o}ra}},
  \bibinfo{author}{\bibfnamefont{F.}~\bibnamefont{Simon}}, \bibnamefont{and}
  \bibinfo{author}{\bibfnamefont{R.}~\bibnamefont{Moessner}},
  \bibinfo{journal}{physica status solidi (RRL)-Rapid Research Letters}
  \textbf{\bibinfo{volume}{7}}, \bibinfo{pages}{101} (\bibinfo{year}{2013}).

\bibitem[{\citenamefont{Jotzu et~al.}(2014)\citenamefont{Jotzu, Messer,
  Desbuquois, Lebrat, Uehlinger, Greif, and Esslinger}}]{jotzu2014experimental}
\bibinfo{author}{\bibfnamefont{G.}~\bibnamefont{Jotzu}},
  \bibinfo{author}{\bibfnamefont{M.}~\bibnamefont{Messer}},
  \bibinfo{author}{\bibfnamefont{R.}~\bibnamefont{Desbuquois}},
  \bibinfo{author}{\bibfnamefont{M.}~\bibnamefont{Lebrat}},
  \bibinfo{author}{\bibfnamefont{T.}~\bibnamefont{Uehlinger}},
  \bibinfo{author}{\bibfnamefont{D.}~\bibnamefont{Greif}}, \bibnamefont{and}
  \bibinfo{author}{\bibfnamefont{T.}~\bibnamefont{Esslinger}},
  \bibinfo{journal}{Nature} \textbf{\bibinfo{volume}{515}},
  \bibinfo{pages}{237} (\bibinfo{year}{2014}).

\bibitem[{\citenamefont{Aidelsburger et~al.}(2015)\citenamefont{Aidelsburger,
  Lohse, Schweizer, Atala, Barreiro, Nascimbene, Cooper, Bloch, and
  Goldman}}]{aidelsburger2015measuring}
\bibinfo{author}{\bibfnamefont{M.}~\bibnamefont{Aidelsburger}},
  \bibinfo{author}{\bibfnamefont{M.}~\bibnamefont{Lohse}},
  \bibinfo{author}{\bibfnamefont{C.}~\bibnamefont{Schweizer}},
  \bibinfo{author}{\bibfnamefont{M.}~\bibnamefont{Atala}},
  \bibinfo{author}{\bibfnamefont{J.~T.} \bibnamefont{Barreiro}},
  \bibinfo{author}{\bibfnamefont{S.}~\bibnamefont{Nascimbene}},
  \bibinfo{author}{\bibfnamefont{N.}~\bibnamefont{Cooper}},
  \bibinfo{author}{\bibfnamefont{I.}~\bibnamefont{Bloch}}, \bibnamefont{and}
  \bibinfo{author}{\bibfnamefont{N.}~\bibnamefont{Goldman}},
  \bibinfo{journal}{Nature Physics} \textbf{\bibinfo{volume}{11}},
  \bibinfo{pages}{162} (\bibinfo{year}{2015}).

\bibitem[{\citenamefont{Rechtsman et~al.}(2013)\citenamefont{Rechtsman, Zeuner,
  Plotnik, Lumer, Podolsky, Dreisow, Nolte, Segev, and
  Szameit}}]{rechtsman2013photonic}
\bibinfo{author}{\bibfnamefont{M.~C.} \bibnamefont{Rechtsman}},
  \bibinfo{author}{\bibfnamefont{J.~M.} \bibnamefont{Zeuner}},
  \bibinfo{author}{\bibfnamefont{Y.}~\bibnamefont{Plotnik}},
  \bibinfo{author}{\bibfnamefont{Y.}~\bibnamefont{Lumer}},
  \bibinfo{author}{\bibfnamefont{D.}~\bibnamefont{Podolsky}},
  \bibinfo{author}{\bibfnamefont{F.}~\bibnamefont{Dreisow}},
  \bibinfo{author}{\bibfnamefont{S.}~\bibnamefont{Nolte}},
  \bibinfo{author}{\bibfnamefont{M.}~\bibnamefont{Segev}}, \bibnamefont{and}
  \bibinfo{author}{\bibfnamefont{A.}~\bibnamefont{Szameit}},
  \bibinfo{journal}{Nature} \textbf{\bibinfo{volume}{496}},
  \bibinfo{pages}{196} (\bibinfo{year}{2013}).

\bibitem[{\citenamefont{Rahav et~al.}(2003)\citenamefont{Rahav, Gilary, and
  Fishman}}]{Fishman}
\bibinfo{author}{\bibfnamefont{S.}~\bibnamefont{Rahav}},
  \bibinfo{author}{\bibfnamefont{I.}~\bibnamefont{Gilary}}, \bibnamefont{and}
  \bibinfo{author}{\bibfnamefont{S.}~\bibnamefont{Fishman}},
  \bibinfo{journal}{Physical Review A} \textbf{\bibinfo{volume}{68}},
  \bibinfo{pages}{013820} (\bibinfo{year}{2003}).

\bibitem[{\citenamefont{Hafezi et~al.}(2011)\citenamefont{Hafezi, Demler,
  Lukin, and Taylor}}]{Hafezi2011Robust}
\bibinfo{author}{\bibfnamefont{M.}~\bibnamefont{Hafezi}},
  \bibinfo{author}{\bibfnamefont{E.~A.} \bibnamefont{Demler}},
  \bibinfo{author}{\bibfnamefont{M.~D.} \bibnamefont{Lukin}}, \bibnamefont{and}
  \bibinfo{author}{\bibfnamefont{J.~M.} \bibnamefont{Taylor}},
  \bibinfo{journal}{Nature Physics} \textbf{\bibinfo{volume}{7}},
  \bibinfo{pages}{907} (\bibinfo{year}{2011}).

\bibitem[{\citenamefont{Hafezi et~al.}(2013)\citenamefont{Hafezi, Mittal, Fan,
  Migdall, and Taylor}}]{Hafezi2013Imaging}
\bibinfo{author}{\bibfnamefont{M.}~\bibnamefont{Hafezi}},
  \bibinfo{author}{\bibfnamefont{S.}~\bibnamefont{Mittal}},
  \bibinfo{author}{\bibfnamefont{J.}~\bibnamefont{Fan}},
  \bibinfo{author}{\bibfnamefont{A.}~\bibnamefont{Migdall}}, \bibnamefont{and}
  \bibinfo{author}{\bibfnamefont{J.}~\bibnamefont{Taylor}},
  \bibinfo{journal}{Nature Photonics} \textbf{\bibinfo{volume}{7}},
  \bibinfo{pages}{1001} (\bibinfo{year}{2013}).

\bibitem[{\citenamefont{Mittal et~al.}(2014)\citenamefont{Mittal, Fan, Faez,
  Migdall, Taylor, and Hafezi}}]{Hafezi2014synthetic}
\bibinfo{author}{\bibfnamefont{S.}~\bibnamefont{Mittal}},
  \bibinfo{author}{\bibfnamefont{J.}~\bibnamefont{Fan}},
  \bibinfo{author}{\bibfnamefont{S.}~\bibnamefont{Faez}},
  \bibinfo{author}{\bibfnamefont{A.}~\bibnamefont{Migdall}},
  \bibinfo{author}{\bibfnamefont{J.}~\bibnamefont{Taylor}}, \bibnamefont{and}
  \bibinfo{author}{\bibfnamefont{M.}~\bibnamefont{Hafezi}},
  \bibinfo{journal}{Physical Review Letters} \textbf{\bibinfo{volume}{113}},
  \bibinfo{pages}{087403} (\bibinfo{year}{2014}).

\bibitem[{\citenamefont{Skirlo et~al.}(2015)\citenamefont{Skirlo, Lu, Igarashi,
  Yan, Joannopoulos, and Solja{\v{c}}i{\'c}}}]{soljacicChern}
\bibinfo{author}{\bibfnamefont{S.~A.} \bibnamefont{Skirlo}},
  \bibinfo{author}{\bibfnamefont{L.}~\bibnamefont{Lu}},
  \bibinfo{author}{\bibfnamefont{Y.}~\bibnamefont{Igarashi}},
  \bibinfo{author}{\bibfnamefont{Q.}~\bibnamefont{Yan}},
  \bibinfo{author}{\bibfnamefont{J.}~\bibnamefont{Joannopoulos}},
  \bibnamefont{and}
  \bibinfo{author}{\bibfnamefont{M.}~\bibnamefont{Solja{\v{c}}i{\'c}}},
  \bibinfo{journal}{Physical Review Letters} \textbf{\bibinfo{volume}{115}},
  \bibinfo{pages}{253901} (\bibinfo{year}{2015}).

\bibitem[{\citenamefont{Quelle and Morais~Smith}(2014)}]{Smith1}
\bibinfo{author}{\bibfnamefont{A.}~\bibnamefont{Quelle}} \bibnamefont{and}
  \bibinfo{author}{\bibfnamefont{C.}~\bibnamefont{Morais~Smith}},
  \bibinfo{journal}{Physical Review B} \textbf{\bibinfo{volume}{90}},
  \bibinfo{pages}{195137} (\bibinfo{year}{2014}).

\bibitem[{\citenamefont{Quelle et~al.}(2016)\citenamefont{Quelle, Goerbig, and
  Morais~Smith}}]{Smith2}
\bibinfo{author}{\bibfnamefont{A.}~\bibnamefont{Quelle}},
  \bibinfo{author}{\bibfnamefont{M.}~\bibnamefont{Goerbig}}, \bibnamefont{and}
  \bibinfo{author}{\bibfnamefont{C.}~\bibnamefont{Morais~Smith}},
  \bibinfo{journal}{New Journal of Physics} \textbf{\bibinfo{volume}{18}},
  \bibinfo{pages}{015006} (\bibinfo{year}{2016}).

\bibitem[{\citenamefont{Kitagawa et~al.}(2012)\citenamefont{Kitagawa, Broome,
  Fedrizzi, Rudner, Berg, Kassal, Aspuru-Guzik, Demler, and
  White}}]{Kitagawabound}
\bibinfo{author}{\bibfnamefont{T.}~\bibnamefont{Kitagawa}},
  \bibinfo{author}{\bibfnamefont{M.~A.} \bibnamefont{Broome}},
  \bibinfo{author}{\bibfnamefont{A.}~\bibnamefont{Fedrizzi}},
  \bibinfo{author}{\bibfnamefont{M.~S.} \bibnamefont{Rudner}},
  \bibinfo{author}{\bibfnamefont{E.}~\bibnamefont{Berg}},
  \bibinfo{author}{\bibfnamefont{I.}~\bibnamefont{Kassal}},
  \bibinfo{author}{\bibfnamefont{A.}~\bibnamefont{Aspuru-Guzik}},
  \bibinfo{author}{\bibfnamefont{E.}~\bibnamefont{Demler}}, \bibnamefont{and}
  \bibinfo{author}{\bibfnamefont{A.~G.} \bibnamefont{White}},
  \bibinfo{journal}{Nature Communications} \textbf{\bibinfo{volume}{3}},
  \bibinfo{pages}{882} (\bibinfo{year}{2012}).

\bibitem[{\citenamefont{Gao et~al.}(2015)\citenamefont{Gao, Gao, Shi, Yang,
  Lin, Xu, Joannopoulos, Solja\v{c}i\'{c}, Chen, Lu
  et~al.}}]{soljacicanomalous}
\bibinfo{author}{\bibfnamefont{F.}~\bibnamefont{Gao}},
  \bibinfo{author}{\bibfnamefont{Z.}~\bibnamefont{Gao}},
  \bibinfo{author}{\bibfnamefont{X.}~\bibnamefont{Shi}},
  \bibinfo{author}{\bibfnamefont{Z.}~\bibnamefont{Yang}},
  \bibinfo{author}{\bibfnamefont{X.}~\bibnamefont{Lin}},
  \bibinfo{author}{\bibfnamefont{H.}~\bibnamefont{Xu}},
  \bibinfo{author}{\bibfnamefont{J.~D.} \bibnamefont{Joannopoulos}},
  \bibinfo{author}{\bibfnamefont{M.}~\bibnamefont{Solja\v{c}i\'{c}}},
  \bibinfo{author}{\bibfnamefont{H.}~\bibnamefont{Chen}},
  \bibinfo{author}{\bibfnamefont{L.}~\bibnamefont{Lu}}, \bibnamefont{et~al.},
  \bibinfo{journal}{arXiv:1504.07809v2}  (\bibinfo{year}{2015}).

\bibitem[{\citenamefont{Hu et~al.}(2015)\citenamefont{Hu, Pillay, Wu, Pasek,
  Shum, and Chong}}]{microwave}
\bibinfo{author}{\bibfnamefont{W.}~\bibnamefont{Hu}},
  \bibinfo{author}{\bibfnamefont{J.~C.} \bibnamefont{Pillay}},
  \bibinfo{author}{\bibfnamefont{K.}~\bibnamefont{Wu}},
  \bibinfo{author}{\bibfnamefont{M.}~\bibnamefont{Pasek}},
  \bibinfo{author}{\bibfnamefont{P.~P.} \bibnamefont{Shum}}, \bibnamefont{and}
  \bibinfo{author}{\bibfnamefont{Y.~D.} \bibnamefont{Chong}},
  \bibinfo{journal}{Physical Review X} \textbf{\bibinfo{volume}{5}},
  \bibinfo{pages}{011012} (\bibinfo{year}{2015}).

\bibitem[{\citenamefont{Garanovich et~al.}(2012)\citenamefont{Garanovich,
  Longhi, Sukhorukov, and Kivshar}}]{garanovich2012light}
\bibinfo{author}{\bibfnamefont{I.~L.} \bibnamefont{Garanovich}},
  \bibinfo{author}{\bibfnamefont{S.}~\bibnamefont{Longhi}},
  \bibinfo{author}{\bibfnamefont{A.~A.} \bibnamefont{Sukhorukov}},
  \bibnamefont{and} \bibinfo{author}{\bibfnamefont{Y.~S.}
  \bibnamefont{Kivshar}}, \bibinfo{journal}{Physics Reports}
  \textbf{\bibinfo{volume}{518}}, \bibinfo{pages}{1} (\bibinfo{year}{2012}).

\bibitem[{\citenamefont{Hatsugai}(1993{\natexlab{a}})}]{bulkedge}
\bibinfo{author}{\bibfnamefont{Y.}~\bibnamefont{Hatsugai}},
  \bibinfo{journal}{Physical Review Letters} \textbf{\bibinfo{volume}{71}},
  \bibinfo{pages}{3697} (\bibinfo{year}{1993}{\natexlab{a}}).

\bibitem[{\citenamefont{Hatsugai}(1993{\natexlab{b}})}]{bulkedge1}
\bibinfo{author}{\bibfnamefont{Y.}~\bibnamefont{Hatsugai}},
  \bibinfo{journal}{Physical Review B} \textbf{\bibinfo{volume}{48}},
  \bibinfo{pages}{11851} (\bibinfo{year}{1993}{\natexlab{b}}).

\bibitem[{\citenamefont{Mukherjee et~al.}(2015)\citenamefont{Mukherjee,
  Spracklen, Choudhury, Goldman, {\"O}hberg, Andersson, and
  Thomson}}]{mukherjee2015observation}
\bibinfo{author}{\bibfnamefont{S.}~\bibnamefont{Mukherjee}},
  \bibinfo{author}{\bibfnamefont{A.}~\bibnamefont{Spracklen}},
  \bibinfo{author}{\bibfnamefont{D.}~\bibnamefont{Choudhury}},
  \bibinfo{author}{\bibfnamefont{N.}~\bibnamefont{Goldman}},
  \bibinfo{author}{\bibfnamefont{P.}~\bibnamefont{{\"O}hberg}},
  \bibinfo{author}{\bibfnamefont{E.}~\bibnamefont{Andersson}},
  \bibnamefont{and} \bibinfo{author}{\bibfnamefont{R.~R.}
  \bibnamefont{Thomson}}, \bibinfo{journal}{Physical Review Letters}
  \textbf{\bibinfo{volume}{114}}, \bibinfo{pages}{245504}
  (\bibinfo{year}{2015}).

\bibitem[{\citenamefont{Davis et~al.}(1996)\citenamefont{Davis, Miura,
  Sugimoto, and Hirao}}]{davis1996writing}
\bibinfo{author}{\bibfnamefont{K.~M.} \bibnamefont{Davis}},
  \bibinfo{author}{\bibfnamefont{K.}~\bibnamefont{Miura}},
  \bibinfo{author}{\bibfnamefont{N.}~\bibnamefont{Sugimoto}}, \bibnamefont{and}
  \bibinfo{author}{\bibfnamefont{K.}~\bibnamefont{Hirao}},
  \bibinfo{journal}{Optics Letters} \textbf{\bibinfo{volume}{21}},
  \bibinfo{pages}{1729} (\bibinfo{year}{1996}).

\bibitem[{\citenamefont{Ams et~al.}(2005)\citenamefont{Ams, Marshall, Spence,
  and Withford}}]{ams2005slit}
\bibinfo{author}{\bibfnamefont{M.}~\bibnamefont{Ams}},
  \bibinfo{author}{\bibfnamefont{G.}~\bibnamefont{Marshall}},
  \bibinfo{author}{\bibfnamefont{D.}~\bibnamefont{Spence}}, \bibnamefont{and}
  \bibinfo{author}{\bibfnamefont{M.}~\bibnamefont{Withford}},
  \bibinfo{journal}{Optics Express} \textbf{\bibinfo{volume}{13}},
  \bibinfo{pages}{5676} (\bibinfo{year}{2005}).

\bibitem[{\citenamefont{Huang and Haus}(1990)}]{Huang}
\bibinfo{author}{\bibfnamefont{W.}~\bibnamefont{Huang}} \bibnamefont{and}
  \bibinfo{author}{\bibfnamefont{H.~A.} \bibnamefont{Haus}},
  \bibinfo{journal}{Journal of Lightwave Technology}
  \textbf{\bibinfo{volume}{8}}, \bibinfo{pages}{922} (\bibinfo{year}{1990}).

\bibitem[{\citenamefont{Horn and Johnson}(1990)}]{Matrix}
\bibinfo{author}{\bibfnamefont{R.}~\bibnamefont{Horn}} \bibnamefont{and}
  \bibinfo{author}{\bibfnamefont{C.}~\bibnamefont{Johnson}},
  \emph{\bibinfo{title}{Matrix Analysis}} (\bibinfo{publisher}{Cambridge
  University Press}, \bibinfo{year}{1990}).

\bibitem[{\citenamefont{Bianco and Resta}(2011)}]{Resta1}
\bibinfo{author}{\bibfnamefont{R.}~\bibnamefont{Bianco}} \bibnamefont{and}
  \bibinfo{author}{\bibfnamefont{R.}~\bibnamefont{Resta}},
  \bibinfo{journal}{Physical Review B} \textbf{\bibinfo{volume}{84}},
  \bibinfo{pages}{241106} (\bibinfo{year}{2011}).

\bibitem[{\citenamefont{Tran et~al.}(2015)\citenamefont{Tran, Dauphin, Goldman,
  and Gaspard}}]{DucThanh}
\bibinfo{author}{\bibfnamefont{D.-T.} \bibnamefont{Tran}},
  \bibinfo{author}{\bibfnamefont{A.}~\bibnamefont{Dauphin}},
  \bibinfo{author}{\bibfnamefont{N.}~\bibnamefont{Goldman}}, \bibnamefont{and}
  \bibinfo{author}{\bibfnamefont{P.}~\bibnamefont{Gaspard}},
  \bibinfo{journal}{Physical Review B} \textbf{\bibinfo{volume}{91}},
  \bibinfo{pages}{085125} (\bibinfo{year}{2015}).

\end{thebibliography}


\newcommand{\beginsupplement}{%
        \setcounter{equation}{0}
        \renewcommand{\theequation}{S\arabic{equation}}%
        \setcounter{figure}{0}
        \renewcommand{\thefigure}{S\arabic{figure}}%
     }
\beginsupplement

\section*{SUPPLEMENTARY MATERIAL}

In this supplementary material we first discuss the fabrication and characterization techniques (Section I). In Section II, we show that two synchronously bending waveguides are equivalent to two straight waveguides with some effective coupling constant. In Section III the techniques used to obtain the topological phase diagram of Fig.~\ref{fig2} (main text) are presented. We discuss wavelength tuning and the effect of disorder in Section IV before finally concluding with a study of how this disorder affects the topology in Section V.

\subsection*{I. Fabrication and characterisation}

The photonic lattice with two driving periods was fabricated inside a 70-mm-long glass (Corning Eagle$^{2000}$) substrate using the ultrafast laser inscription technique, where the refractive index profile of each waveguide was controlled using the ``slit-beam shaping" method \cite{davis1996writing, ams2005slit}. The glass substrate, mounted on $x$-$y$-$z$ translation stages, was translated at 8~mm/s through the focus of a 500~kHz train of 1030~nm femtosecond laser pulses to fabricate each waveguide. The laser inscription parameters were optimized to produce waveguides that were single-mode and well confined in the measurement wavelength range of 700-830~nm. To study the response of the edge modes in the presence of a defect at the edge, the waveguide at the (8, 4) lattice site was not fabricated. It should also be highlighted that the waveguide paths are designed such that all waveguides exhibit identical bend radii at a given $z$, although the direction of this bending is site dependent. This ensures that there is  minimal site-dependent losses.

To measure  the coupling constants, we fabricated five sets of each bond separately inside the same substrate. These bonds (or couplers) were then characterised in the wavelength range 705-795~nm, to obtain the variation of the mean and standard deviation of coupling strength with wavelength (see supplementary for details).
To excite the lattices with different wavelengths, a photonic crystal fiber was pumped by sub-picosecond laser pulses of 1064~nm wavelength to generate a broadband supercontinuum. A tunable monochromator placed after the supercontinuum source was used to select narrow band ($\approx 3$~nm) light, which was coupled into an optical fibre (SMF-600). The fiber was then coupled to the lattice sites. The output intensity distribution was observed using a CMOS camera. A polarizer passing only vertically polarized light is placed in front of the camera to ensure that the measurements are not affected by polarization-dependent coupling.

\subsection*{II. Coupling between bending waveguides}
In the experimental setup, the lattice constant was chosen such that, in the absence of any bending, the waveguides comprising the lattice were practically uncoupled. In order to turn on the coupling between two neighbouring sites [see Fig.~\ref{fig1}~(b) in the main text], the two waveguides were bent together and then remained straight before moving apart again [see Fig.~\ref{fig1}~(d) in the main text].  The coupling between two such synchronously bending waveguides was demonstrated in Ref.~\cite{Huang} to be equivalent to an effective coupling between two straight waveguides. This mapping is vital in order for the experiment to emulate our theoretical model, which assumes that the lattice is comprised of straight waveguides but with controllable couplings to neighbouring waveguides [see Fig.~\ref{fig1}~(b) in the main text]. In light of the importance of this result in the present work, a derivation is provided here for completeness.
The dynamics of how light propagates through an array of weakly-guiding waveguides is governed by the scalar Helmholtz equation,
\begin{equation}
    \nabla^2\psi+\frac{1}{\lambdabar^2}n^2(x,y,z)\psi=0. \notag
\end{equation}
Here $\psi(x,y,z)$ is the electric field amplitude, $\lambdabar=\frac{\lambda}{2\pi}$, where $\lambda$ is the wavelength of the light in free space, and $n(x,y,z)$ is the refractive index profile in the domain of interest. If the wave propagation is primarily along the $z$-axis, then the field $E(x,y,z)$ can be represented as a slowly varying complex field amplitude and a fast oscillating wave, $\psi(x,y,z)=E(x,y,z)e^{i\frac{n_s}{\lambdabar}z}$. Substituting $\psi$ into the Helmholtz equation and making the slowly varying envelope approximation for $E$ gives
\begin{equation}
     i\lambdabar\frac{\partial E}{\partial z}=-\frac{\lambdabar^2}{2n_s}(\frac{\partial^2}{\partial x^2}+\frac{\partial^2}{\partial x^2})E+\frac{1}{2n_s}(n_s^2-n^2)E.
     \label{Para}
\end{equation}
The refractive index profile can be written as $n(x,y,z)=n_s+\Delta n(x,y,z)$. However, in laser-written waveguide arrays, the refractive index change caused by the laser is small and so $\frac{1}{2n_s}(n_s^2-n^2)\approx -\Delta n(x,y,z)=V[x,y,z]$.
Consider now an array of moving waveguides where $V[x,y,z]$ is given by
\begin{equation}
V[x,y,z]=\sum_k V_0^{k}
=\sum_{k} V_0[x-x_{k}^0-x_{k}(z),y-y_{k}^0].
\notag
\end{equation}
Here $V_0(x,y)$ is the refractive index profile of a single isolated waveguide, the index $k$ labels the different waveguides and $(x_k^{0}+x_k(z),y_k^{0})$ is the $z$-dependent position of the $k$th waveguide.  \\
In the regime where the waveguides are far apart, compared to the width of $V_0$, the electric field can be well approximated by a superposition of the fundamental modes of the different waveguides,
\begin{equation}
E(x,y,z)=\sum_k c_k(z)u[x-x_k(z),y]e^{-i\beta z},
\notag
\end{equation}
where $u[x,y]$ is the fundamental mode of an isolated waveguide and satisfies the equation
\begin{equation}
    -\beta u[x,y]=-\frac{\lambdabar^2}{2n_s}(\frac{\partial^2}{\partial x^2}+\frac{\partial^2}{\partial y^2})u[x,y]+V_0[x,y]u[x,y].
    \label{Eigenmode}
\end{equation}
Inserting the electric field expansion into Eq.~\eqref{Para} gives, after some rearrangement,
\begin{align}
&i\lambdabar\sum_k(\dot{c}_k u_k+c_k\frac{\partial u_k}{\partial x}(-\dot{x}_k))\!=\!\sum_k c_k \bigg [\!-\!\frac{\lambdabar^2}{2n_s}\bigg(\frac{\partial^2}{\partial x^2}\!+\!\frac{\partial^2}{\partial y^2}\bigg)\nonumber \\
&+\beta +V_0^{k} \bigg]u_k+ \sum_{n \neq k} V_0^{n} \sum_k c_k u_k,
\end{align}
where we have introduced the notation $u_k=u[x-x_k^0-x_k(z),y-y_k^0]$. The expression in the square brackets is equal to zero as a result of Eq.~\eqref{Eigenmode}.
\\  Multiplying by $u_m$ and integrating transversely gives
\begin{equation}
i\sum_k(\dot{c}_k p_{m,k}+c_k f_{m,k})=\sum_k c_k t_{m,k},
\label{CMT}
\end{equation}
where
\begin{eqnarray}
\notag
p_{m,k}=& \int_{-\infty}^{\infty}\int_{-\infty}^{\infty} dxdy u_m u_k,\\
\notag
f_{m,k}=& -\dot{x}_k \int_{-\infty}^{\infty}\int_{-\infty}^{\infty} dxdy u_m \frac{\partial u_k}{\partial x},\\
\notag
t_{m,k}=& \frac{1}{\lambdabar}\sum_{n\neq k} \int_{-\infty}^{\infty}\int_{-\infty}^{\infty} dxdy  V_0^{n} u_m u_k.
\end{eqnarray}
A number of pertinent statements can be made about these integrals:  
\begin{eqnarray}
\notag
p_{m,k}=&p_{k,m},\\
\notag
f_{m,m}=&0,\\
\notag
f_{m,k}=&-\dot{x}_k I_{m,k},\\
\notag
f_{k,m}=&\dot{x}_m I_{m,k},\\
\notag
I_{m,k}=& \int_{-\infty}^{\infty}\int_{-\infty}^{\infty} dxdy u_m \frac{\partial u_k}{\partial x}.
\notag
\end{eqnarray}
These statements have the corollary that
\begin{equation}
    \frac{\partial }{\partial z} p_{m,k}=\dot{p}_{m,k}=I_{m,k}(\dot{x}_{m}-\dot{x}_k).
    \notag
\end{equation}
 As previously mentioned, in the absence of any bending there is negligible coupling between any of the waveguides. The consequence of this is that in the presence of bending, only pairs of waveguides are coupled at any given moment. The coupled mode equations for this system can then be written as block diagonal matrices with each block comprising a $2\times 2$ matrix. This simplifies the following to a study of a two waveguide coupler as the analysis will straightforwardly generalise to the full lattice. Therefore, if there are two waveguides, labelled $1$ and $2$, that are synchronously bending, such that $\dot{x}_1=-\dot{x}_2$, then it can be readily seen that $\dot{p}_{1,2}=2f_{1,2}=2f_{2,1}$. The coupled-mode equations for this $2 \times 2$ block can therefore be written in matrix notation as
\begin{equation}
    i(\textbf{P}\dot{\textbf{C}}+\frac{1}{2}\dot{\textbf{P}}\textbf{C})=\textbf{T}\textbf{C},
    \label{MatForm}
\end{equation}
where the matrices $\textbf{P}$ and $\textbf{T}$ have the form
\begin{equation}
\textbf{P}=
\begin{pmatrix}
1 &  X(z) \\
X(z)  &  1\\
\end{pmatrix}; \quad \textbf{T}=
\begin{pmatrix}
0 &  \kappa_{12}(z) \\
\kappa_{12}(z) &  0\\
\label{PMat}
\end{pmatrix}.
\end{equation}
Here $X(z)=p_{1,2}(z)$ and $\kappa_{12}(z)=t_{1,2}(z)$. 
 To proceed further we introduce a new set of variables by $\textbf{C}=\textbf{M}\textbf{W}$, where $\textbf{M}$ is chosen such that $\textbf{M}^{\dagger}\textbf{P}\textbf{M}=\textbf{I}$. This transformation is always possible as both $\textbf{I}$ and $\textbf{P}$ are both Hermitian, positive definite matrices and so can be connected via a similarity transformation by a non-singular matrix~\cite{Matrix}. This change of variable matrix can be written as
 \begin{equation}
\textbf{M}=\frac{1}{\sqrt{2}}
\begin{pmatrix}
\frac{1}{\sqrt{1+X(z)}} &  -\frac{1}{\sqrt{1-X(z)}} \\
\frac{1}{\sqrt{1+X(z)}}  &  \frac{1}{\sqrt{1-X(z)}}\\
\end{pmatrix}.
\label{MMat}
\end{equation}
Using this change of variables and multiplying from the left by $\textbf{M}^{\dagger}$ allows Eq.~\eqref{MatForm} to be rewritten as
\begin{equation}
i\dot{\textbf{W}}= \textbf{M}^{\dagger}\textbf{T}\textbf{M}\textbf{W},
\notag
\end{equation}
with $\textbf{M}^{\dagger}\textbf{T}\textbf{M}$ taking the diagonal form
\begin{equation}
\textbf{M}^{\dagger}\textbf{T}\textbf{M}=
\begin{pmatrix}
\frac{\kappa_{12}(z)}{1+X(z)} &  0 \\
0  & \frac{\kappa_{12}(z)}{-1+X(z)} \\
\end{pmatrix}.
\notag
\end{equation}
The equation for $\dot{\textbf{W}}$ can be straightforwardly solved to yield 
\begin{equation}
    \textbf{W}(z)=\textbf{T}(z)\textbf{W}_0,\\
    \notag
\end{equation}
where $\textbf{T}$ has the form
\begin{align}
&\textbf{T}(z)=\begin{pmatrix}
\int_0^{z}dz \frac{\kappa_{12}(z)}{1+X(z)} & 0 \\
0  & \int_0^{z}dz \frac{\kappa_{12}(z)}{-1+X(z)}\\
\end{pmatrix}\nonumber\\
&\approx \begin{pmatrix}
\int_0^{z}dz \kappa_{12}(z) & 0 \\
0  & -\int_0^{z}dz \kappa_{12}(z)\\
\end{pmatrix}.
\notag
\end{align} 
In the second equality, the cross-power term, $X(z)$, has been neglected, which is a valid approximation as long as the waveguides remain far apart throughout the bending motion. \\
A return to the original $\textbf{C}$ variables can be made by inverting the non-singular matrix $\textbf{M}$,
\begin{equation}
    \textbf{C}(z)= \textbf{M}(z)\textbf{T}(z)\textbf{M}^{-1}(0)\textbf{C}_0,
    \notag
\end{equation}
where $\textbf{C}_0$ are the initial conditions.

The waveguides in the experimental lattice are coming together and then moving apart.
Consequently,  at the end of the bending section, $z=L$, the waveguides are at the same separation as at $z=0$ and so $\textbf{M}(L)=\textbf{M}(0)$. This allows the matrix product $\textbf{M}(L)\textbf{U}(L)\textbf{M}^{-1}(0)$ to take the form 
\begin{equation}
\textbf{M}(L)\textbf{T}(L)\textbf{M}^{-1}(0)=\textbf{U}(L)=
\begin{pmatrix}
\cos(\phi) &  -i\sin(\phi) \\
-i\sin(\phi)  & \cos(\phi) \\
\end{pmatrix},
\label{Evo}
\end{equation}
with
\begin{equation}
\phi=\int_0^{L} dz \kappa_{12}(z).
\notag
\end{equation}
For comparison, the evolution operator for two straight waveguides of length $L$ is given by
\begin{equation}
\textbf{U}_{Str}(L)=
\begin{pmatrix}
\cos(JL) &  -i\sin(JL) \\
-i\sin(JL)  & \cos(JL) \\
\end{pmatrix},
\notag
\end{equation}
where $J$ is the coupling between the waveguides.
As can be readily observed, this has the same form as Eq.~\eqref{Evo} if we write  $\phi$ as $\phi=LJ_\text{eff}$ where $J_\text{eff}$ is an effective coupling. 

This analysis has been conducted for two waveguides ,but can be readily extended to the full lattice discussed in the main text. The experimental lattice is written such that only pairs of waveguides are coupled at any given moment. The $\textbf{P}$ and $\textbf{M}$ matrices therefore have block-diagonal matrix form with the blocks all having the form of Eq.~\eqref{PMat}.
\subsection*{III. Method to compute the topological phase diagram}
\begin{figure}[h!]
\includegraphics[width=8.5cm]{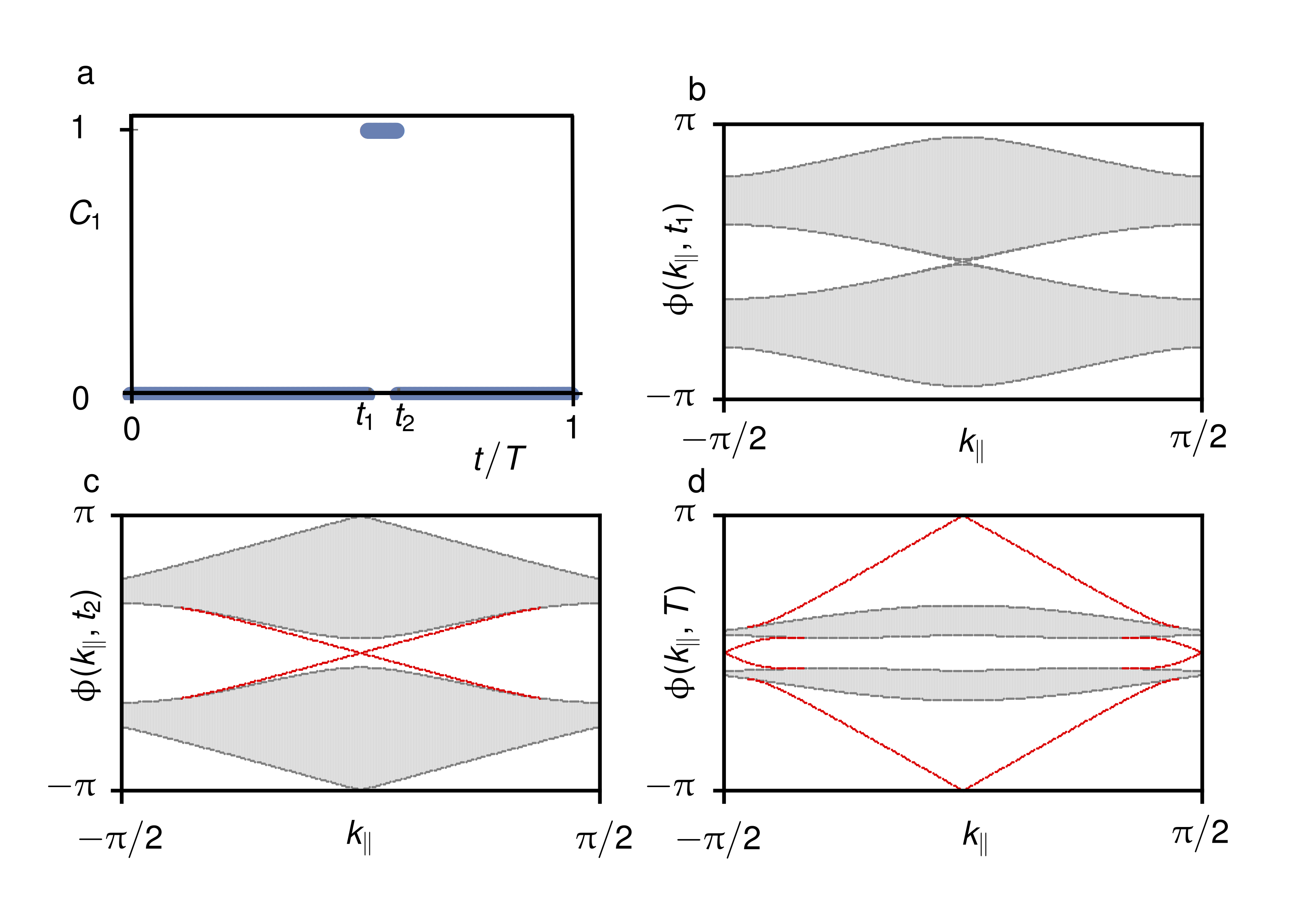}
\caption{{\bf Example calculation of the winding number using the evolution of the instantaneous Chern number.} (a) Evolution of the instantaneous Chern number of the lowest band, $C_1$ over one driving period; the system parameters are $\Lambda_1\!=\!1$ and $\Lambda_2\!=\!1.4$. The Chern number changes twice within a driving period (at times $t_1$ and $t_2$). The first of these topological transitions occurs when the gap closes at $\phi\!=\!0$, (b), whilst  the second occurs at $\phi\!=\!\pi$, (c). This changing of the instantaneous Chern number that occurs through the zone edge is responsible for the winding numbers $W_1$ and $W_2$ being non-zero, and the Chern number of the Floquet bands to be zero. This non-trivial value for the winding number  allows for the presence of edge modes in the spectrum at $t=T$, (d), even though the Chern numbers of all the Floquet bands (evaluated at $t\!=\!T$) are zero.}
\label{Instant}
\end{figure}

In this section the methodology used to obtain the topological phase diagram in Fig.~$\ref{fig2}$ (main text) is detailed; this method is based on Ref.~\cite{nathan2015topological}, and we refer the reader to this reference for more details on concepts related to the topological characterisation of Floquet systems. The calculation of the different phases is assisted by utilizing the property that the topology of the system can only change when there is a gap closing between the two bulk bands. The position of these gap-closing events can be found analytically by diagonalizing the evolution operator at the end of the driving period. It is found that for $\Lambda_1\!=\!\Lambda_2$ and $\Lambda_2\!=\!\frac{1}{3}(2\pi-\Lambda_1)$ the system is gapless at quasienergy zero, whilst for $\Lambda_2=\frac{1}{3}(\pi-\Lambda_1)$ and  $\Lambda_2=\frac{1}{3}(3\pi-\Lambda_1)$ the system is gapless through the fundamental zone edge. The position of these gap closings thereby divides the phase space into the eight different sectors shown in Fig.~$\ref{fig2}$. The topology within these different sectors can then be defined by calculation of the winding numbers.  As discussed in Eq.~\eqref{Proj}, at any time $t$ the evolution operator may be diagonalized to yield the instantaneous Bloch bands of the driven system. The eigenstates associated with these bands can be used to calculate an instantaneous Chern number for that band. In accordance with Eq.~\eqref{Winding} the winding number can then be calculated by tracking changes in the Chern number of the lowest band that occur through the zone edge throughout the driving period. This procedure is illustrated in Fig.~$\ref{Instant}$ (a) for the parameters $\Lambda_1=1$, $\Lambda_2=1.4$, which shows how the instantaneous Chern number of the lowest band changes throughout the driving period. It can be readily observed that that the Chern number changes twice within a driving period and computing the spectra at these times shows that the second of these changes occurs through the zone edge, see Fig.~$\ref{Instant}$~(b-c). This latter degeneracy causes the Chern number of the lowest band to decrease by one. This topological transition through the zone edge when combined with the Chern numbers of all the bands being zero at $t=T$  implies, in accordance with  Eq.~\eqref{Winding}, that both $W_1$ and $W_2$ are equal to one.
\subsection*{IV. Wavelength Tuning and Disorder}
\begin{figure}[h!]
\includegraphics[width=8.5cm]{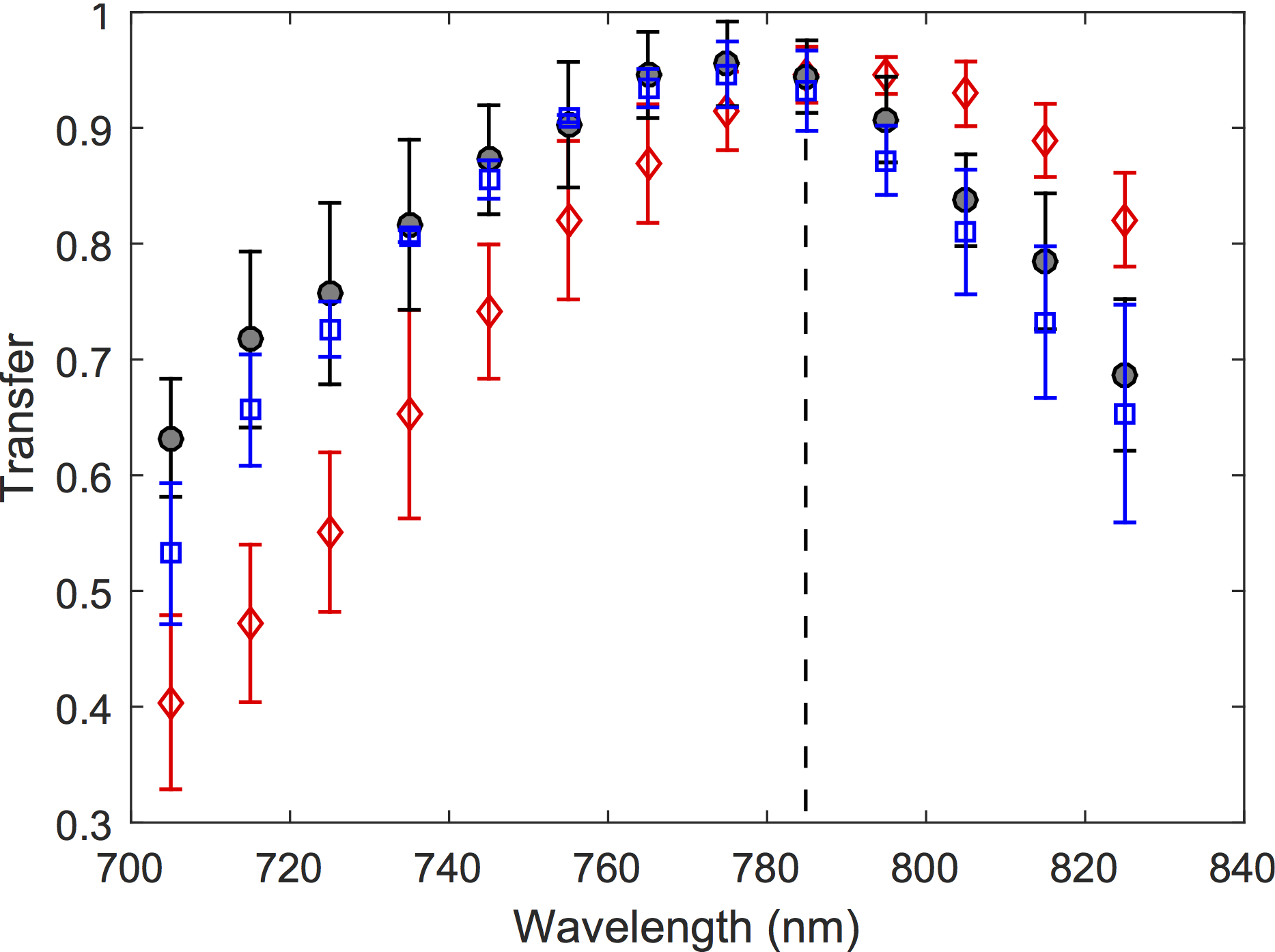}
\caption{{\bf{Variation of transfer for bonds $J_{2-4}$ as a function of wavelength.}}
The effective coupling can be extracted from this characterisation data using the relation Transfer $=\sin^2(J_\text{eff}L)$, where 
$L$ is the physical length of the sample. The dotted line indicates that the transfer is very close to 100$\%$ for all the three bonds at 785~nm wavelength.}
\label{Coupling}
\end{figure}

In the main text it was discussed how the experimental setup provides  the ability to modify the wavelength of light used as an input. This ability allows the effective coupling between waveguides to be altered without modifying the lattice. In order to understand the behaviour of the bonds $J_{2-4}$, see main text, as a function of wavelength, five sets of isolated bonds were written and the percentage of light transferred from the launch waveguide as a function of wavelength was measured, Fig.~\ref{Coupling}.  This bond characterisation data illustrates how at 785~nm the bonds $J_{2-4}$ have almost equal transfer which is close to 100$\%$. This wavelength therefore closely matches the desired parameters discussed in the main text of having three bonds equal and perfectly transferring.

\begin{figure}[h!]
\includegraphics[width=8.6cm]{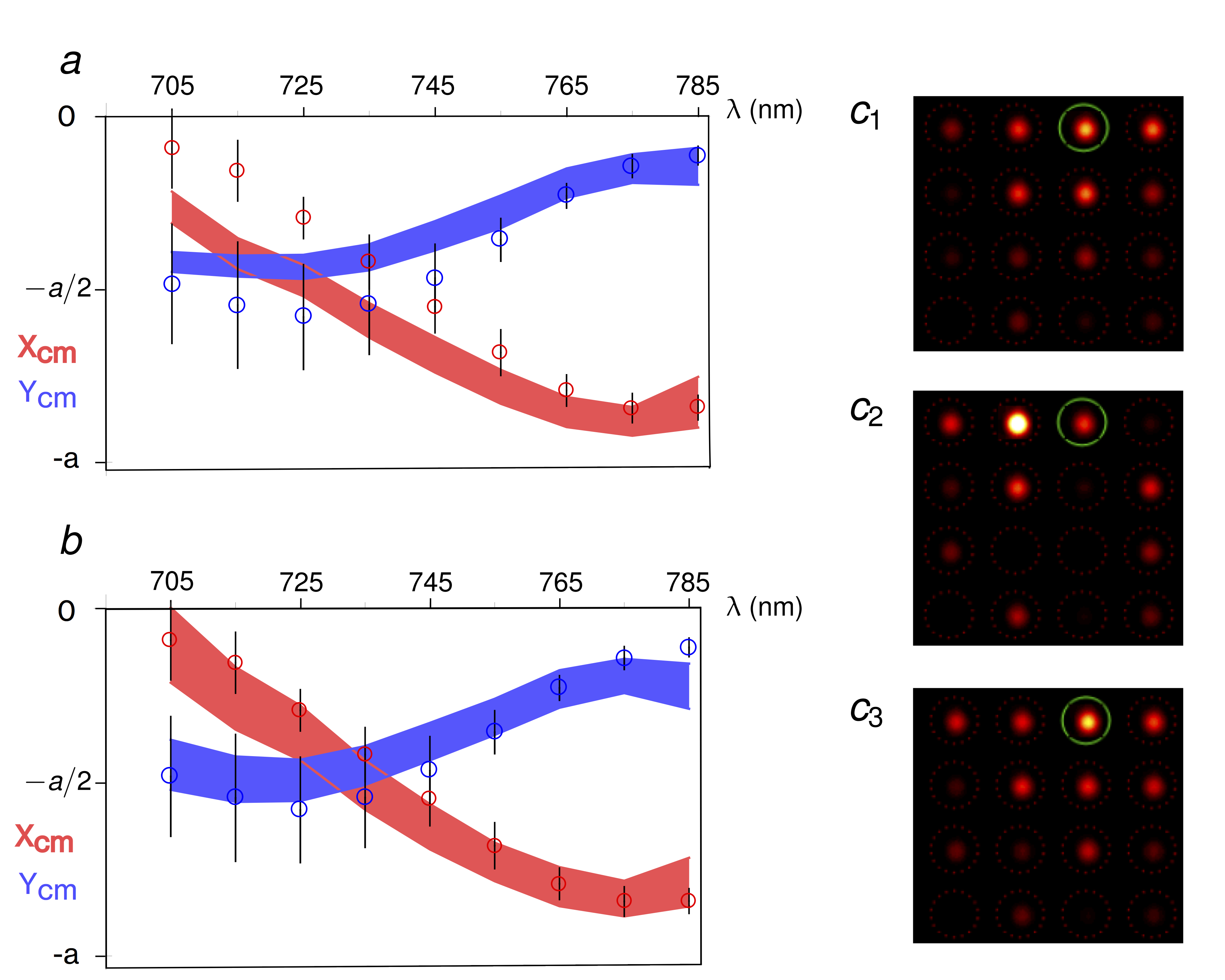}
\caption{\textbf{Comparing experimental results and theoretical predictions from two different disorder models.} Model 1 has bond strength disorder only whilst model 2 has additional phase disorder. (a) and (b) compare the experimental centre of mass drifts, as defined by Eq.~\eqref{EqnCM}, to those calculated theoretically for models 1 and 2, respectively. The centre of mass drift for four different launch sites is measured experimentally and the red and blue circles indicate the average drift in the $x$ and $y$ directions whilst the black error bars indicate the range of values measured experimentally. The shaded regions in (a) and (b) indicate the theoretical disorder averaged value plus/minus one standard deviation. (C$_1$) is the experimental output image averaged over the 4 launch sites for 705~nm input whilst (C$_2$) and (C$_3$) are the disordered averaged output facet images coming from models 1 and 2 respectively.}
\label{CM}
\end{figure}

The capability of this wavelength tuning technique in our experimental setup allows for the transfer of the bonds to be changed substantially from $\approx100 \%$ right down to $\approx50\%$. This provides the ability to compare experimental and theoretical predictions for a wide range of parameters. In order to perform such a comparison, the coupling strengths extracted from the bond characterisation data were used to calculate a theoretical centre of mass drift, $\textbf{r}_{cm}$, that could be compared to experimental results. The centre of mass drift is defined as
\begin{equation}
    \textbf{r}_{cm}=\sum_{m,n} I_{m,n}\textbf{r}_{m,n}-\textbf{r}_0,
    \label{EqnCM}
\end{equation}
where $I_{m,n}$ is the intensity in the $(m, n)$-th site at the output, $\textbf{r}_{m,n}$ is the coordinates of the $(m, n)$-th  site and $\textbf{r}_0$ are the coordinates of the launch site. Experimental data was taken at four different physical locations in the lattice. These different positions should, in the absence of disorder, produce identical results for $\textbf{r}_{cm}$, up to a sign change. However, the non-zero standard deviations measured in the bond characterisation data indicates the presence of bond strength disorder within the lattice (such a disorder will be referred to as ``off-diagonal disorder"). The effects of this disorder can be directly observed in the experimental lattice with the four different launch sites producing slightly different results for $\textbf{r}_{cm}$, see the error bars in  Fig.~\ref{CM}(a). In order to theoretically model this disorder, the strengths of the bonds $J_{2-4}$ were randomly selected to lie within the range measured in the bond characterisation data. The comparison between the
experimental data and the theoretical prediction  shows good agreement around $\lambda\!\approx\!785$~nm, but deviations are observed at smaller wavelengths, Fig.~\ref{CM}(a). The failure of the theoretical model in this low transfer region is further illustrated in Fig.~\ref{CM}(C$_{1-2}$), which compares the averaged experimental output image to the theoretical image averaged over 1000 disorder realisations.  

The match between theory and experimental results can be improved by including a small on-site disorder term, $\Delta \beta_{m,n}$, in addition to the coupling disorder of the previous model (this form of disorder will be referred to as ``diagonal disorder"). The $\Delta \beta_{m,n}$ for the $(m,n)$-{th} site is a random number drawn from within a uniform distribution covering the interval $[-W,\!\quad\!W]$. The parameter $W$ is the disorder strength which is chosen as $WT/4=0.6$, as this particular value results in the best fit to the experimental data over the whole wavelength range investigated.  The corresponding center-of-mass prediction, Fig.~\ref{CM}(b), and output facet image, Fig.~\ref{CM}(C$_{3}$), produced by this model are in closer agreement to the experimental results which points towards a disordering of this type being present in the lattice. 

\subsection*{V. Study of Disorder and its impact on topological bands}

The analysis conducted in the previous section illustrated the presence of off-diagonal disorder as well as possibly indicating the presence of diagonal disorder terms.  In the absence of any disorder, the experimental lattice should reproduce a driven-lattice model that exhibits anomalous topological edge modes (i.e.~chiral edge modes that are associated with non-zero winding numbers, while the Chern number of the Floquet bands is trivial; see main text).  The inclusion of disorder in the system can allow bandgap closings to occur which can modify the topology of the system. In particular, the disorder could potentially change the Chern numbers of the Floquet bands to a non-trivial value such that the experimentally observed edge modes would no longer be ``anomalous". The elimination of this possibility thereby requires the calculation of the Chern number in a disordered system.   

The presence of disorder means that quasimomentum is no longer a good quantum number which in turn precludes the use of the usual momentum space techniques in calculating the Chern number. In order to overcome this limitation we use a real-space Chern invariant, $C(\epsilon_0)$, as introduced by Bianco and Resta in Ref. \cite{Resta1}. This invariant allows for a generalisation of the ({\textit{static}}) bulk-edge correspondence to disordered and quasi-periodic lattices~\cite{Resta1,DucThanh}. In particular, when the reference energy, $\epsilon_0$, is placed in a mobility gap of the spectrum the Chern invariant will measure the number of chiral edge modes which cross this energy, as predicted by the Chern numbers of the bands below this energy.  This real-space Chern invariant is based around an operator called the Chern marker, $\hat{\mathcal{C}}$, which is defined by
\begin{equation}
\hat{\mathcal{C}}=-4\pi \text{Im}[\hat{x}_Q\hat{y}_P].
\end{equation}
Here the operators $\hat{\textbf{r}}_{P}=\hat{P}\hat{\textbf{r}}\hat{Q}$ and $\hat{\textbf{r}}_{Q}=\hat{Q}\hat{\textbf{r}}\hat{P}$
are expressed in terms of the position operator $\hat{\textbf{r}}=(\hat{x},\hat{y})$ and the projection operators
\begin{equation}
\hat{P}(\epsilon_0)=\sum_{\epsilon\leq \epsilon_0} |\psi_{\epsilon} \rangle \langle \psi_{\epsilon}|=\hat{1}-\hat{Q}.
\end{equation}
In the absence of disorder a real space Chern Number can be defined by averaging the Chern marker operator over a unit cell,
\begin{equation}
C=\int_{cell} \langle \textbf{r}| \hat{\mathcal{C}}|\textbf{r}\rangle d\textbf{r}=\int_{cell} \mathcal{C}(\textbf{r}) d\textbf{r}.
\end{equation}

\begin{figure}[t!]
\includegraphics[width=8.6cm]{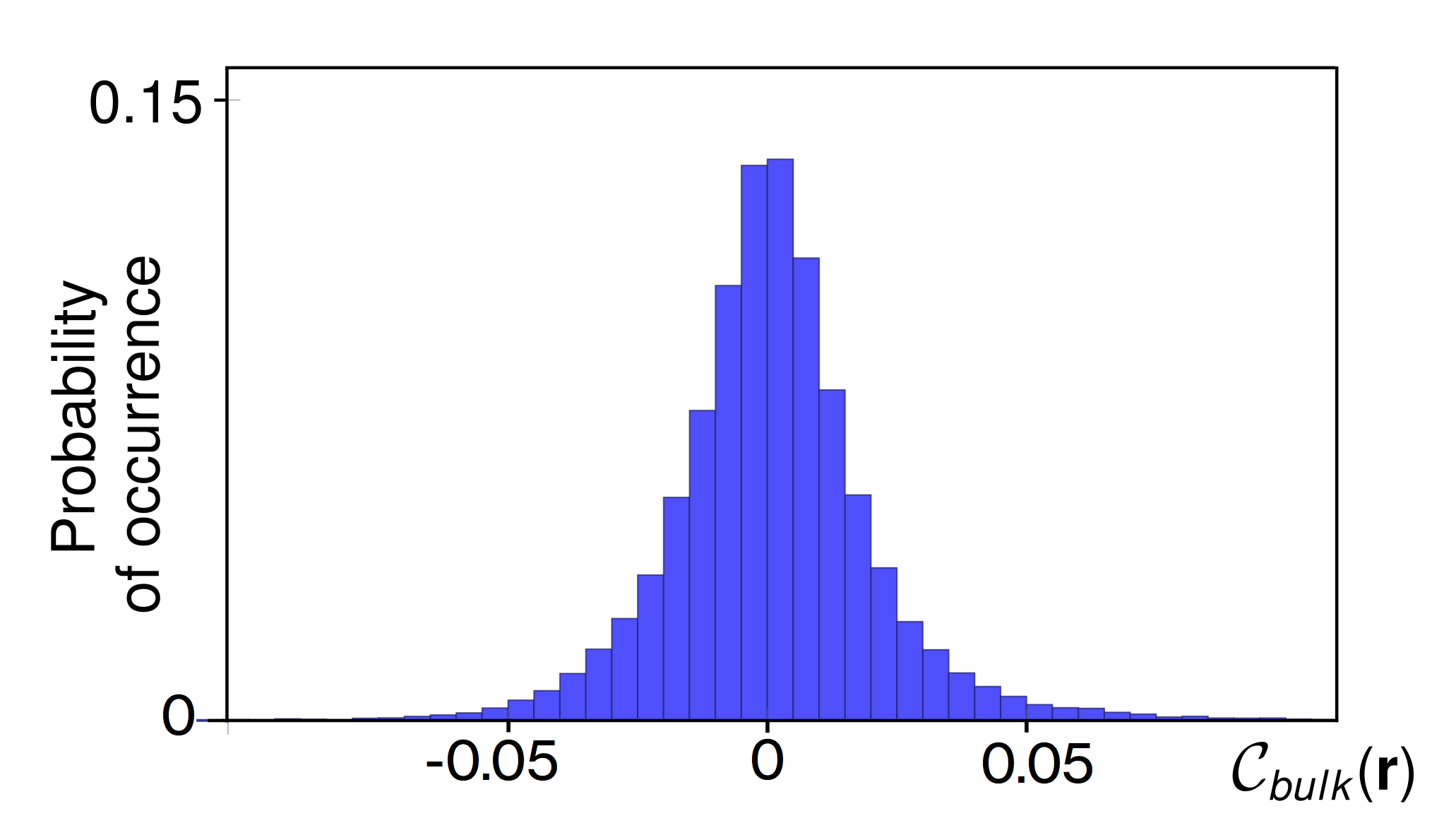}
\caption{\textbf {The local Chern marker for the experimental lattice with the addition of both diagonal and off-diagonal disorder terms.} The figure shows the probability of occurrence of a particular value for $\mathcal{C}(\textbf{r})$ anywhere within the bulk of the lattice. This local Chern marker is narrowly peaked around zero which implies that the static bulk-edge correspondence would predict no robust chiral edge modes present in the system.}
\label{Chern}
\end{figure}

In the presence of disorder this real-space Chern number fluctuates depending upon the position of the unit cell. These fluctuations can be accounted for by replacing the unit-cell average by an average over an area $A$ that is located within the bulk and is large compared to the fluctuation length~\cite{DucThanh}. 

These ideas thereby provide a mechanism to investigate whether the disorder terms discussed in the previous section can alter the Chern numbers of the Floquet bands to non-trivial values.  In particular we calculate the local Chern marker, in the presence of these disorder terms, for a reference energy of zero, $\epsilon_0=0$. The real-space average of this local Chern marker, the Chern invariant, can be viewed as the prediction coming from the static bulk-edge correspondence for the number of chiral edge modes that are traversing the mobility gap between the two bands.  The results, however, of this analysis reveal that even in the presence of disorder the local Chern marker, $\mathcal{C}(\textbf{r},0)$, is still heavily peaked around zero with this local quantity fluctuating only slightly within the bulk, Fig.~$\ref{Chern}$. The Chern invariant is therefore equal to zero independently of the size of the averaging area that is chosen. Consequently, the static bulk-edge correspondence  predicts the experimental lattice should not feature any chiral edge modes.

\end{document}